\documentclass[twocolumn]{autart}    

\usepackage{graphicx}         
\usepackage{amssymb}
\usepackage{amsmath}
\usepackage{float}
\usepackage{algorithm,algorithmic}
\usepackage{subcaption}
\usepackage{mathtools}
\usepackage{tcolorbox}
\newtheorem{example}{Example}
\newtheorem{theorem}{Theorem}
\newtheorem{corollary}[theorem]{Corollary}
\newtheorem{definition}[theorem]{Definition}
\usepackage{tikz}
\usetikzlibrary{positioning,arrows.meta}
\begin{document}

\begin{frontmatter}

\title{A Game-Theoretic Framework for Distributed Load Balancing: Static and Dynamic Game Models} 

\thanks[footnoteinfo]{This work was supported by the Air Force Office of Scientific Research under grant FA9550-23-1-0107.}
\author[fatemeh]{Fatemeh Fardno}\ead{ffardno2@illinois.edu},    
\author[S. rasoul]{S. Rasoul Etesami}\ead{etesami1@illinois.edu}               

\thanks[fatemeh]{Fatemeh Fardno is with the Department of Electrical and Computer Engineering and Coordinated Science Laboratory, University of Illinois Urbana-Champaign, Urbana, IL 61801.}
\thanks[S. rasoul]{ Rasoul Etesami is with the Department of Industrial and Systems Engineering, Department of Electrical and Computer Engineering, and Coordinated Science Laboratory, University of Illinois Urbana-Champaign, Urbana, IL 61801.}


\begin{abstract}                          
Motivated by applications in job scheduling, queuing networks, and load balancing in cyber-physical systems, we develop and analyze a game-theoretic framework to balance the load among servers in static and dynamic settings. In these applications, jobs/tasks are held by selfish entities that do not want to coordinate with each other, yet the goal is to balance the load among servers in a distributed manner. First, we provide a static game formulation in which each player holds a job with a specific processing requirement and wants to schedule it fractionally among a set of heterogeneous servers to minimize its average processing time. We show that this static game is a potential game with a pure Nash equilibrium (NE). In particular, the best-response dynamics converge to such an NE after $n$ iterations, where $n$ is the number of players. Additionally, we bound the price of anarchy (PoA) of the static game in terms of game parameters. We then extend our results to a dynamic game setting, where jobs arrive and get processed, and players observe the load on the servers to decide how to schedule their jobs. In this setting, we show that if the players update their strategies using dynamic best-response, the system eventually becomes fully load-balanced and the players' strategies converge to the pure NE of the static game. In particular, we show that the convergence time scales only polynomially with respect to the game parameters. Finally, we provide numerical results to evaluate the performance of our proposed algorithms.  
\end{abstract}

\end{frontmatter}

\section{Introduction}
\label{sec:introduction}
Distributed and uncoordinated load balancing is challenging when a set of strategic agents compete for a set of shared resources. There is a large body of literature on load balancing, in which the proposed schemes can be either static or dynamic. In static load balancing schemes \cite{berenbrink2007distributed,grosu2005noncooperative,grosu2002load,schaerf1994adaptive}, agents use prior knowledge about the system, whereas dynamic load balancing methods \cite{abel2025learning,gaitonde2020stability,gaitonde2021virtues,kameda1997comparison,krishnasamy2016regret} make decisions based on the system's real-time state.

Load balancing in strategic environments has been studied using game theory in the existing literature, where, in most applications, the system is often modeled as congestion games \cite{milchtaich1996congestion,holzman1997strong}. In congestion games, each player's utility is determined by the resources they select and the number of other players choosing the same resources. If all players choose the most beneficial resource, it becomes congested, and all players suffer from the congestion cost. As a result, agents may prefer to choose less profitable resources to avoid congestion. 

One of the limitations of the aforementioned models is the assumption that rounds are independent, i.e., the outcome of one round does not affect future rounds \cite{gaitonde2020stability}. However, in queuing systems, jobs left in queues create a carryover effect, making each round’s average wait time dependent not only on newly assigned jobs but also on residual jobs. Motivated by these limitations and applications in job scheduling and queuing networks, we propose and analyze a game-theoretic framework for load balancing in both static and dynamic settings. Here, jobs are managed by self-interested players who do not coordinate, yet the goal is to achieve decentralized and balanced server loads.

We first introduce a static load balancing game model in which each player holds a job with a certain processing requirement and seeks to schedule it fractionally across a set of heterogeneous servers to minimize their average processing time. Next, to account for the carryover effect, we extend our model to a dynamic game setting, where jobs arrive and are processed in the system over time. In this setting, players observe the load (state) on the servers and decide how to schedule their jobs to minimize their cumulative average processing time. We also assume that, at each time step, only one player receives a job. In both settings, we show that if players update their strategies using static or dynamic best-response at each state, the system eventually becomes load-balanced, and the players' strategies converge to the pure Nash equilibrium (NE) of the underlying game. Notably, we show that the convergence time scales polynomially with respect to the game parameters in both settings, which is further justified using extensive numerical experiments.

\subsection{Related Work}
Load balancing in distributed systems has been extensively studied in the literature. Many of these studies focus on systems with a centralized scheduler \cite{zhou2018degree,hellemans2018power,horvath2019mean,mitzenmacher2001power,lin1992dynamic}. Among the various load balancing approaches, join-the-shortest-queue (JSQ) stands out as one of the most thoroughly researched methods \cite{gupta2007analysis,eschenfeldt2018join,foley2001join,kogias2019r2p2,kogias2020hovercraft,gardner2019smart,winston1977optimality}. This approach assigns jobs to the server with the shortest queue length. Join-idle-queue (JIQ) \cite{lu2011join,mitzenmacher2016analyzing,jennings2015resource} is another load balancing policy where jobs are sent to idle servers, if any, or to a randomly selected server. These methods might not perform well because the queue length is not always a good indicator of the servers' wait times. Power-of-\(d\)-choices \cite{mitzenmacher2001power,xie2015power,gardner2017redundancy,ousterhout2013sparrow,roy2015chaos,nasir2015power,zhu2020racksched} queries \(d\) randomly selected servers and sends the jobs to the least loaded ones. This method has been proven to yield short queues in homogeneous systems but may be unstable in heterogeneous systems.

Studying distributed load balancing through a game-theoretic lens is a natural approach. Despite its potential, relatively few studies have applied game theory to the load balancing problem \cite{kameda2012optimal,roughgarden2001stackelberg,nash1950bargaining,economides1991multi,economides1990game,orda1993competitive,altman2001routing,korilis1997capacity,koutsoupias1999worst,mavronicolas2001price,roughgarden2002bad}. In \cite{grosu2005noncooperative}, the authors model the load balancing problem in heterogeneous systems as a non-cooperative, repeated finite game and identify a structure for the best-response policy in pure strategies. They later introduce a distributed greedy algorithm to implement the best-response policy. However, since their setting is stateless, the proposed algorithm does not necessarily perform well in dynamic games. Additionally, \cite{grosu2002load} models the load balancing problem as a cooperative game among players. Unlike distributed load balancing algorithms without queue memory \cite{grosu2005noncooperative,grosu2002load}, which rely on stateless updates, our model explicitly captures the state dynamics and shows that stability and convergence can be guaranteed in polynomial time, even under strategic behavior of the players.

The work \cite{schaerf1994adaptive} studies the distributed load balancing problem using multi-agent reinforcement learning methods. In this approach, each server simultaneously works on multiple jobs, and the efficiency with which the server handles a job depends on its capacity and the number of jobs it processes. The problem is modeled as a stochastic system where each player has access to local information. To evaluate server efficiency, each player maintains an efficiency estimator, a vector summarizing the server's performance, which also serves as the player's state. The players' policies are then formulated as functions of these states. They compare the performance of their algorithm to other existing algorithms, such as non-adaptive algorithms, i.e., algorithms where players use servers in a predetermined manner, while revealing that naive communication between players does not necessarily improve the system's overall efficiency.

Krishnasamy et al. \cite{krishnasamy2016regret} study a variant of the multi-armed bandit problem with the carryover effect, which is suitable for queuing applications. They consider a bandit setting in which jobs queue for service, and the service rates of the different queues are initially unknown. The authors examine a queuing system with a single queue and multiple servers, where both job arrivals to the queue and the service rates of the servers follow a Bernoulli distribution. At each time slot, only one server can serve the queue, and the objective is to determine the optimal server scheduling policy for each time slot. They further evaluate the performance of different scheduling policies against the \emph{no-regret policy}\footnote{The policy that schedules the optimal server at each time.}.
\par 

More recently, \cite{gaitonde2020stability} studied the multi-agent version of the queuing system in \cite{krishnasamy2016regret}. Their setting differs from ours in the sense that, in our setting, each server has a separate queue and processes jobs with a fixed service rate. However, \cite{gaitonde2020stability} describes a system where each player has a queue, and servers process jobs with a fixed time-independent probability. All unprocessed jobs are then sent back to their respective queues. Queues also receive bandit feedback on whether their job was cleared by their chosen server. One concern in such settings is whether the system remains stable, despite the selfish behavior of the learning players. This concern does not arise in settings with a centralized scheduler, as these systems remain stable if the sum of arrival rates is less than the sum of service rates. However, this condition does not guarantee stability in settings with strategic players. In \cite{gaitonde2020stability}, the authors show that if the capacity of the servers is high enough to allow a centralized coordinator to complete all jobs even with double the job arrival rate, and players use \emph{no-regret} learning algorithms,\footnote{A no-regret learning algorithm minimizes the expected difference in queue sizes relative to a strategy that knows the optimal server.} then the system remains stable. They later show in \cite{gaitonde2021virtues} that stability can be guaranteed even with an extra capacity of \( \frac{e}{e-1}\), where $e\approx 2.718$ is the natural number, but only if players choose strategies that maximize their long-run success rate. Subsequently, \cite{abel2025learning} extends the model in \cite{gaitonde2020stability} by allowing servers to hold a single packet for later service if it cannot be served immediately, and by removing the need to prioritize older packets. Compared to no-regret learning in queuing systems \cite{abel2025learning,gaitonde2020stability,gaitonde2021virtues}, which emphasizes learning behavior under adversarial and stochastic arrival processes, our framework adopts a strategic, game-theoretic approach, where agents optimize best responses against evolving queue states. We establish polynomial-time convergence under a deterministic update rule, rather than relying on regret minimization.

Our dynamic game fundamentally differs from repeated congestion games with discounting \cite{krichene2015online,tennenholtz2009learning}, which assume that each round is independent of the others. First, in repeated homogeneous congestion games, all players sharing a resource receive the same reward, which decreases with increased usage \cite{rosenthal1973class}. In contrast, our model allows heterogeneity, where players using the same resource may receive different rewards. Second, we incorporate a carryover effect: queue states evolve across rounds based on their current state and the number of players choosing them. Rather than optimizing discounted payoffs, we focus on long-run system behavior and establish convergence guarantees based on average cost over time.

Finally, our work is also related to \cite{cui2022learning}, where the authors introduce a new class of \emph{Markov congestion games}, which model non-stationarity in congestion games. They consider the carryover effect, and at each time step~$t$ and state~$s$, the players participate in a static congestion game. However, the authors show that even with a centralized algorithm, computing a NE remains computationally intractable. Our setting differs in two key aspects: we assume asynchronous policy updates for players and deterministic state transitions. Moreover, in our model, players using the same resource can receive different rewards. Under these assumptions, we show that our algorithm converges to a load-balanced equilibrium state in polynomial time.

\subsection{Contributions}

This paper introduces a game-theoretic framework to address the problem of uncoordinated load balancing in systems with self-interested players. The key contributions of this work are as follows:

\begin{itemize}
    \item \textbf{Static Game Model:} We formulate a static game in which each player holds one job with a specific processing requirement and aims to schedule it fractionally across a set of servers to minimize its average wait time. We show that this static game is a potential game and, therefore, admits a pure NE. Moreover, we show that best-response dynamics converge to this pure NE in a number of iterations equal to the number of players. Additionally, we bound the price of anarchy (PoA) of the static game in terms of game parameters, such as service rates, processing requirements, and initial loads.
    
    \item \textbf{Dynamic Game Extension:} To account for the carryover effect in queuing systems, we extend the static model to a dynamic game setting, where at each time step, only one player receives a job. In this extension, jobs arrive and are processed over time, and players update their scheduling strategies based on the current state of the servers, which is fully observable to them. We show that if players use dynamic best-response strategies, the system eventually reaches a load-balanced state, and their strategies converge to the pure NE of the static game. Furthermore, we show that convergence occurs in a polynomial number of iterations. 
    \item \textbf{Numerical Evaluation:} We provide numerical results that evaluate the performance of our proposed algorithms in both static and dynamic settings, demonstrating the efficiency of our approach in achieving load balancing.
\end{itemize}

In summary, our work fills a gap in the literature by deriving a closed-form best response for both the static and dynamic load balancing games, proving single-pass convergence for strategic updates in the static game and bounding its price of anarchy, and establishing polynomial-time convergence for the dynamic game with queue carryover.

\subsection{Organization and Notations}
In Section~\ref{sec:formulation}, we introduce the static load-balancing game model, followed by its extension to a dynamic game setting. Section~\ref{sec:static} presents our main results for the static game model. In Section~\ref{sec:dynamic}, we propose an algorithm for the dynamic game model and analyze its convergence properties. Section~\ref{sec:numerical} provides numerical simulations to evaluate the performance of the proposed algorithms. Finally, Section~\ref{sec:conclusion} concludes the paper and discusses directions for future research. Omitted proofs and other auxiliary results are provided in the Appendix. The main notations used in this paper are listed in Table~\ref{tab:notations}.

\begin{table}[H]
\centering
\small  
\setlength{\tabcolsep}{4pt}
\scalebox{0.9}{
\begin{tabular}{|c|c|}
\hline
\textbf{Notation} & \textbf{Definitions} \\
\hline
$n$ & number of players \\
$m$ & number of servers \\
$\lambda_i$ &  job length of player $i$ \\
$\lambda_{\text{max}}$ & maximum job length across all players\\
$\lambda_{\text{min}}$ & minimum job length across all players\\
$\mu_j$ & service rate of server $j$  \\
$\mu_{\text{max}}$ & maximum service rate across all servers\\
$\mu_{\text{min}}$ & minimum service rate across all servers\\
$a^t_{ij}$ & fraction of job $i$ scheduled on server $j$ at time $t$ \\
$a^t_i$ & action vector of player $i$ at time $t$\\
$s_j^0$ & initial load on server $j$ \\
$s_j^t$ & load on server $j$ at time $t$ \\
\hline
\end{tabular}}
\caption{List of notations}
\label{tab:notations}
\end{table}

\section{Problem Formulation}\label{sec:formulation}
In this section, we first introduce the static load balancing game and then extend it to a dynamic
game setting.
\subsection{Static Load Balancing Game}
\label{sec:problem_formulation_static}
In the static load balancing game, there are a set $[m]=\{1,2,\ldots,m\}$ of $m$ servers with service rates $\mu_j>0$ $\forall j\in [m]$, and a set $[n]=\{1,2,\ldots,n\}$ of $n$ players. Each player $i\in [n]$ holds a job with processing requirement (length) $\lambda_i$. This job can be fractionally scheduled on different servers.\footnote{Since there is a one-to-one correspondence between jobs and players, we often use the terms players and jobs interchangeably.} We also define $\lambda_{\max}=\max_i \lambda_i$ and $\lambda_{\min}=\min_i \lambda_i$ to be the maximum and minimum length of jobs across all the players, respectively.  

\begin{figure}[htbp]
    \centering
    \includegraphics[width=0.4\textwidth]{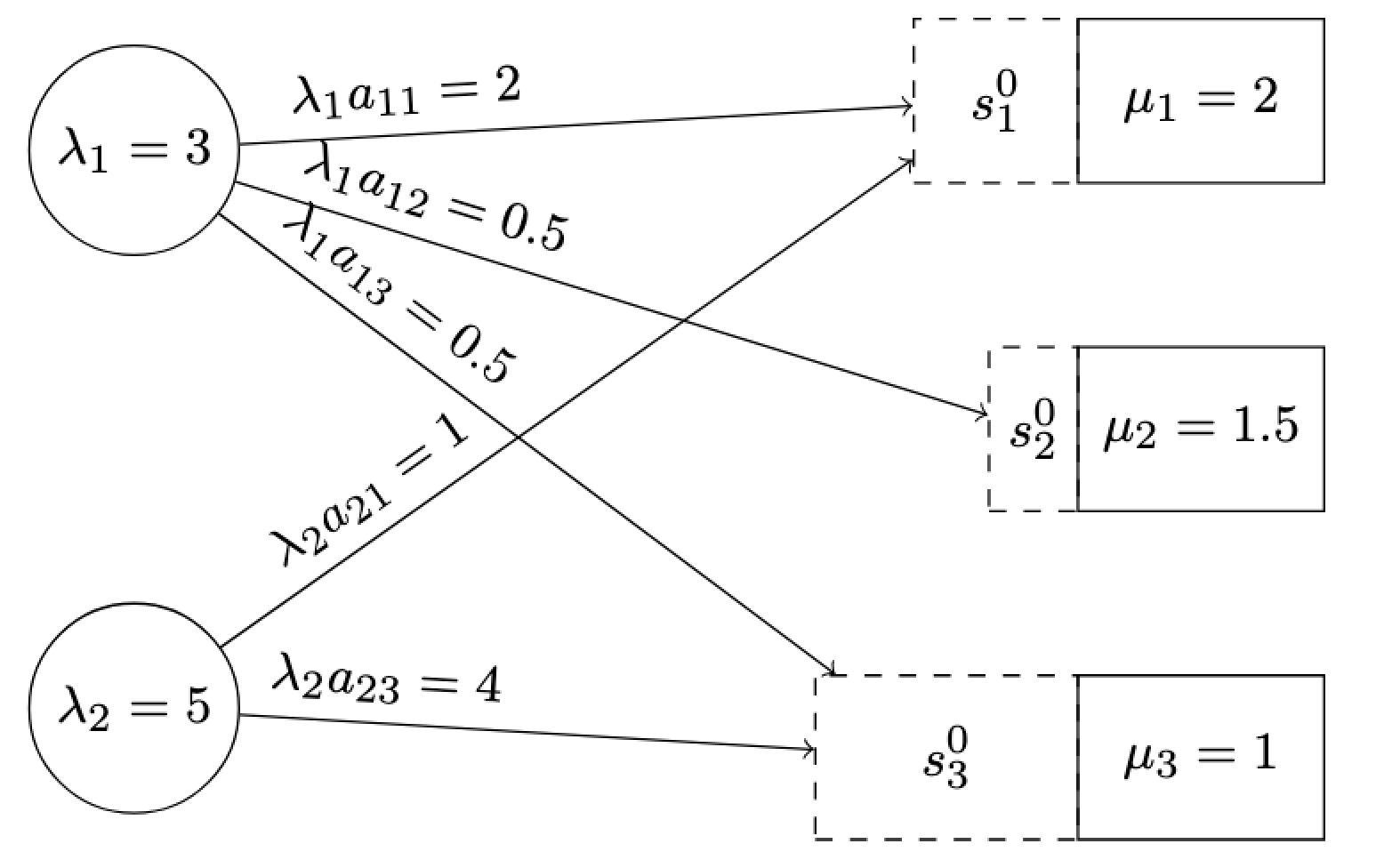} 
    \caption{An illustration of the static load balancing game with two players and three servers.}
    \label{fig:graph}
\end{figure}

Let $a_{ij}$ be the fraction of job $i$ that is scheduled on server $j$ by player $i$. Then, the strategy (action) set for player $i$ is given by all the vectors $a_i=(a_{i1},\ldots,a_{im})$ that belong to the probability simplex
\begin{equation}\nonumber
 A_i=\Big\{a_i: \sum_{j=1}^ma_{ij}=1, a_{ij}\ge 0\Big\}.    
\end{equation}
We further assume that each server $j$ is initially occupied with a load of $s^0_j\ge 0$ jobs. A graphical illustration of the model is provided in Fig.~\ref{fig:graph}. In this example, we consider three servers with initial loads \( s_1^0, s_2^0, s_3^0 \) and service rates \( \mu_1 = 2 \), \( \mu_2 = 1.5 \), and \( \mu_3 = 1\), respectively. Two players are present, with job lengths \( \lambda_1 = 3 \) and \( \lambda_2 = 5 \). As shown in the figure, Player 1 has distributed 2 units of its job on Server 1, 0.5 units on Server 2, and the remaining 0.5 units on Server 3. Meanwhile, Player 2 schedules 1 unit of its job on Server 1 and the remainder on Server 3.

Now, let us suppose player $i$ has already scheduled $x\in [0, \lambda_i)$ portion of its job on server $j$. The amount of time required by server $j$ to fully process that portion is given by $\frac{x+s^0_j}{\mu_j}$. In particular, if player $i$ schedules an additional infinitesimal $dx$ portion of its job on server $j$, the delay experienced by that infinitesimal portion to be processed is $\frac{x+s^0_j}{\mu_j}dx$. Therefore, the wait time for player $i$ when scheduling a fraction $\alpha\in [0, 1]$ of its job on server $j$ can be calculated as:
\begin{equation}\nonumber
    \int_{0}^{\alpha \lambda_i} \frac{x+s_j^0}{\mu_j}  \ dx = \frac{\alpha^2 \lambda_i^2}{2\mu_j} + \frac{\alpha \lambda_i s_j^0}{\mu_j} = \alpha \lambda_i \left(\frac{\alpha \lambda_i}{2\mu_j} + \frac{s_j^0}{\mu_j}\right).    
\end{equation}
Finally, since player $i$ schedules different fractions of its job on different servers according to its action  $a_i$, the \emph{average wait time} of player $i$ when taking action $a_i$ can be written as:
\begin{equation}\label{eq:average-waite}
 \sum_{j=1}^m \lambda_i a_{ij} \left(\frac{\lambda_i a_{ij}}{2\mu_j} + \frac{s_j^0}{\mu_j}\right).
\end{equation}
Using the same reasoning, given an action profile $a=(a_1,\ldots,a_n)$ of all the players, we can define the cost function for player $i$ as 
\begin{equation}\label{eq:cost1-static}
    D_i(a) = \sum_{j=1}^m \lambda_i a_{ij} \left(\frac{\lambda_i a_{ij}}{2\mu_j}+\frac{s^0_j+\sum_{k\neq i}\lambda_k a_{kj}}{\mu_j}\right).
\end{equation}
Therefore, we obtain a noncooperative static load balancing game $([n], \{D_i(a)\}_{i\in n}, A_1\times\cdots\times A_n)$, where each player $i$ aims to make a decision $a_i\in A_i$ in order to minimize its own cost function $D_i(a)$. 

\subsection{Dynamic Load Balancing Game}
\label{sec:problem_formulation_dynamic}

The dynamic game setting is the same as the static one, except that the scheduling process evolves over time as more jobs arrive and get processed. More specifically, in the dynamic load balancing game, at each discrete time $t=0,1,2,\ldots$, we assume that only one player (say player $i$) receives a job\footnote{For instance, if job arrivals follow a continuous random process such as a Poisson process, one may assume that at any time only one of the players receives a new job.}, and this player must immediately schedule it among the servers. Let us define $a_i^t$ and $s_{j}^t$ as the action of player $i$ and the load (state) of server $j$ at time $t$, respectively. Moreover, we denote the \emph{state} of the game at time $t$ by $s^t = (s_{1}^t,\ldots,s^t_m)$, which is the vector of observed loads on each of the servers at time $t$. 
Now suppose that player $i$ receives a new job of length $\lambda_i$ at time $t$ and schedules it among the servers according to its action $a_i^t = (a_{i1}^t, \cdots, a_{im}^t)$. By following the same derivations as in the static game (Eq.~\eqref{eq:average-waite}), the instantaneous average wait time for player $i$ due to its new job arrival at time $t$ equals
\begin{equation}
\label{costFunctionSingaleAgent}
    D_i(a^t,s^t) = \sum_{j=1}^m \lambda_i a^t_{ij} \left( \frac{\lambda_i a^t_{ij}}{2\mu_j}+\frac{s^t_{j}}{\mu_j}\right).
\end{equation}
After that, the state of server $j$ changes according to
\begin{equation}\label{eq:state-change}
    s_{j}^{t+1} = (s_{j}^{t} +  \lambda_i a^t_{ij} - \mu_j)^{+}\ \ \ \forall j\in [m],
\end{equation}
where $(\cdot)^+=\max\{0, \cdot\}$. 

Finally, the goal for each player $i$ is to follow a certain scheduling policy $\pi_i=(a^t_i, t=1,2,\ldots)$ to minimize its overall averaged cumulative cost given by
\begin{equation}\label{eq:cumulative-cost}
V_i(\pi_i,\pi_{-i})=\limsup_{T\to \infty} \frac{1}{T}\sum_{t=1}^{T}D_i(a^t,s^t).    
\end{equation}

\begin{assum}\label{as:stable}
In the dynamic load balancing game, we assume that $\lambda_{\max} < \sum_{j=1}^m \mu_j$.  
\end{assum}

We note that Assumption \ref{as:stable} is the standard stability condition from queuing theory \cite{kleinrock1975queueing} and is crucial in the dynamic setting. Otherwise, one could always consider a worst-case instance where, regardless of any scheduling policy, the length of the waiting jobs in the server queues will grow unbounded.

We conclude this section with an example that illustrates the updates in the dynamic load-balancing game.

\begin{example}\label{example}
    Consider a simple setting with $m = 2$ servers and $n = 2$ players. Suppose player~1's job length is $\lambda_1 = 1$, and player~2's job length is $\lambda_2 = 2$. The initial server loads are $s^0 = (2, 4)$, and the servers' service rates are $\mu_1 = 1.5$ and $\mu_2 = 2.5$. Suppose that at time $t = 0$, player~2 receives a job and decides to split it evenly between the two servers, i.e., $a_{21} = a_{22} = \frac{1}{2}$. An illustrative diagram of the model at time $t = 0$ is shown in Fig.~\ref{fig:dynamic-1}.
    \begin{figure}[htbp]
    \centering
    \includegraphics[width=0.4\textwidth]{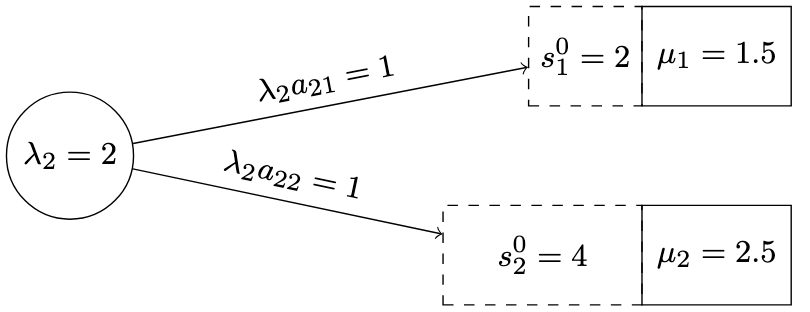} 
    \caption{Illustration of the dynamic game at time $t=0$.}
    \label{fig:dynamic-1}
    \end{figure}
    
    Using \eqref{eq:state-change} we can write
    \begin{align*}
    &s_1^1 = (2+2\left(\frac{1}{2}\right)-1.5)^{+} = 1.5,\cr   
    &s_2^1 = (4+2\left(\frac{1}{2}\right)-2.5)^{+} = 2.5.
    \end{align*}
    Therefore, we have $s^1 = (1.5, 2.5)$, and player 2 incurs an instantaneous cost of
    \begin{align*}
        D_2(a^0, s^0) &= \lambda_2 a^0_{21} \left(\frac{\lambda_2a^0_{21}}{2\mu_1} + \frac{s_1^0}{\mu_1}\right) \\&+ \lambda_2 a^0_{22} \left(\frac{\lambda_2a^0_{22}}{2\mu_2} + \frac{s_2^0}{\mu_2}\right) \\&=
        \left(\frac{1}{3} + \frac{2}{1.5}\right) + \left(\frac{1}{5} + \frac{4}{2.5}\right) \\&=
        3.47.
    \end{align*}    
    Now suppose that at time $t = 1$, player~1 receives a job and sends all of it to server~1. Fig.~\ref{fig:dynamic-2} illustrates the dynamic game at time $t = 1$.
    \begin{figure}[t]
    \centering
    \includegraphics[width=0.4\textwidth]{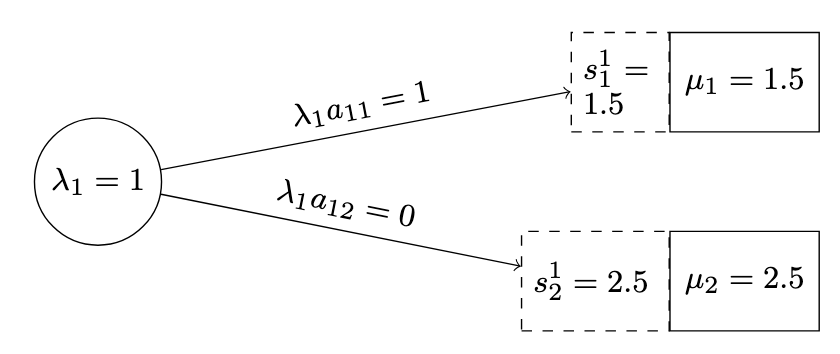} 
    \caption{Illustration of the dynamic game at time $t=1$.}
    \label{fig:dynamic-2}
    \end{figure}
\end{example}

\section{Theoretical Results for the Static Game}\label{sec:static}

In this section, we analyze the static load balancing game and show that it belongs to the class of \emph{exact potential games}, for which a pure NE (defined below) is guaranteed to exist \cite{monderer1996potential}. 
\begin{definition}
An action profile $a=(a_i,a_{-i})$ is called a pure NE for the static load balancing game if 
\begin{align}\nonumber
D_i(a_i,a_{-i})\leq D_i(a'_i,a_{-i})\ \ \forall i, a'_i\in A_i.   
\end{align}
\end{definition}
In fact, in exact potential games, best response dynamics—where players iteratively and sequentially redistribute their jobs across servers by best responding to others' strategies—are known to converge to a pure NE \cite{monderer1996potential}. Thus, we next proceed to characterize the best response dynamics in a closed-form as a solution to a convex program. This closed-form solution will be used in the subsequent section to extend our results to the dynamic game setting. Our main results for the static load balancing game are presented in Lemma~\ref{br-static} and Theorems \ref{static-convergence-NE}, \ref{thm:stateic-n-iteration}, and \ref{thm:PoA}.

\begin{thm}
\label{static-convergence-NE}
    The static load-balancing game is an exact potential game, that is, for any player $i$ and any two action profiles $(a_i, a_{-i})$ and $(a'_i, a_{-i})$ that differ only in the action of player~$i$, we have
\begin{equation}\label{eq:potential-difference}
    \Phi(a_i,a_{-i})-\Phi(a'_i,a_{-i})=D_i(a_i,a_{-i})-D_i(a'_i,a_{-i}).
    \end{equation}In particular, the game admits a pure NE, and the best-response dynamics converge to a pure NE.
\end{thm}
\begin{pf}
 Let us consider the potential function
\begin{equation}\nonumber
\Phi(a) =
    \sum_{j=1}^{m} \frac{\mu_j}{2} \left(\frac{s^0_j+\sum_{k=1}^n\lambda_k a_{kj}}{\mu_j} \right)^2.
\end{equation}

For simplicity, let $s_{ij}=s^0_j+\sum_{k\neq i}\lambda_k a_{kj}$ be the effective load on server $j$ viewed from the perspective of player $i$, where we note that since $s_{ij}, j\in [m]$ do not depend on player $i$'s action, they must be the same for both action profiles $(a_i,a_{-i})$ and $(a'_i,a_{-i})$. We can write    
\begin{align*}
    \Phi(a_i,a_{-i}) =&
    \sum_{j=1}^{m} \frac{\mu_j}{2} \left(\frac{s^0_j+\sum_{k=1}^n\lambda_k a_{kj}}{\mu_j} \right)^2 \\=&
    \sum_{j=1}^{m} \frac{\mu_j}{2} \left( \frac{\lambda_i a_{ij}}{\mu_j} + \frac{s_{ij}}{\mu_j} \right)^2 \\=&
    \sum_{j=1}^{m} \frac{\mu_j}{2} \left(\frac{s_{ij}}{\mu_j} \right)^2 + \sum_{j=1}^{m} \lambda_i a_{ij} \left( \frac{\lambda_i a_{ij}}{2 \mu_j} + \frac{s_{ij}}{\mu_j}\right) \\=&
    \sum_{j=1}^{m} \frac{\mu_j}{2} \left(\frac{s_{ij}}{\mu_j} \right)^2 + D_i(a_i,a_{-i}),
\end{align*}
where the last equality is obtained by the expression of the cost function \eqref{eq:cost1-static} and the definition of $s_{ij}$. Using an identical argument, we have $\Phi(a'_i,a_{-i})=\sum_{j=1}^{m} \frac{\mu_j}{2} \left(\frac{s_{ij}}{\mu_j} \right)^2 + D_i(a'_i,a_{-i})$. Now, by subtracting this relation from the former one, we obtain the desired result in~\eqref{eq:potential-difference}. Finally, it is known from \cite{monderer1996potential} that any exact potential game admits a pure NE, and that any sequence of best-response updates converges to a pure NE.$\hfill{\blacksquare}$
\end{pf}



\begin{rem}
A more intuitive way to see why $\Phi(a)$ serves a potential function is to note that 
\begin{equation}\nonumber
\frac{\partial \Phi(a)}{\partial a_{ij}}=\frac{\partial D_i(a)}{\partial a_{ij}}=\lambda_i \left(\frac{\lambda_i a_{ij} + s_{ij}}{\mu_j} \right)\ \forall i,j.
\end{equation}
This in view of \cite[Lemma 4.4]{monderer1996potential}   suggests that the static load balancing game is an exact potential game.  
\end{rem}

\begin{rem}
One can show that the potential function $\Phi(a)$ in the proof of Theorem~\ref{static-convergence-NE} is strongly convex with respect to each player's action separately. However, in general, $\Phi(a)$ may not be jointly convex with respect to the entire action profile $a$. 
\end{rem}
Next, we proceed to characterize the best response of each player in a closed-form. To that end, let us define $\hat{\mu}_{ij}$ to be the relative \emph{available}  processing rate of server $j$ that is viewed by player $i$, i.e., 
\begin{align*}
\hat{\mu}_{ij} = \frac{\mu_j}{s^0_j+\sum_{k\neq i}\lambda_k a_{kj}}.
\end{align*}
Then, we can rewrite the cost function \eqref{eq:cost1-static} as
\begin{equation}
\label{eq:cost2-static}
    D_i(a) = \sum_{j=1}^m \lambda_i a_{ij} \left( \frac{1}{\hat{\mu}_{ij}} + \frac{\lambda_i a_{ij}}{2\mu_j}\right).
\end{equation}
Given fixed strategies of the other players $a_{-i}$, a simple calculation shows that the entries of the Hessian matrix of the cost function \eqref{eq:cost2-static} with respect to $a_i$ are given by   \begin{align*}        
        \frac{\partial^2 D_i(a)}{\partial a_{ij}\partial a_{i\ell}}=\begin{cases}
       \frac{\lambda_i^2}{\mu_j}\  &\mbox{if}\ \ell=j,\\
        0 \  &\mbox{if}\ \ell\neq j.
        \end{cases}
    \end{align*}
Therefore, the Hessian of $D_i(a)$ with respect to $a_i$ is a diagonal matrix with strictly positive diagonal entries, and hence, it is a positive-definite matrix. This shows that the cost function of each player is strongly convex with respect to its own action.

Now suppose player $i$ plays its best-response action to the fixed actions of the other players (including the initial loads on the servers). As in \cite{grosu2005noncooperative}, we can write a convex program whose solution gives us the best strategy for player $i$:  
\begin{align}
\label{static-convex-programming}
    & \text{minimize} \quad D_i(a) \cr
    & \text{s.t.} \quad\quad \sum_{j=1}^m a_{ij} = 1, \cr
    &\ \quad\qquad a_{ij} \geq 0 \quad \forall j\in [m].
\end{align}
    
Since the objective function $D_i(a)$ is strongly convex with respect to the decision variable $a_i$, we can use KKT optimality conditions to solve the optimization problem \eqref{static-convex-programming} in a closed-form. The result is stated in the following lemma.
\begin{lem}
\label{br-static}
    Given player $i$, assume the servers are sorted according to their available processing rates, i.e.,  $\hat{\mu}_{i1} \geq \hat{\mu}_{i2} \geq \cdots \ge \hat{\mu}_{im}$. Then, the optimal solution of the convex program \eqref{static-convex-programming} is given by
    \begin{align}\label{bestResponsePolicyStatic}
        a_{ij} = 
        \begin{cases} 
            \frac{\mu_j}{\lambda_i} \left( \frac{\lambda_i + \sum_{k=1}^{c_i - 1} \frac{\mu_k}{\hat{\mu}_{ik}}}{\sum_{k=1}^{c_i - 1}\mu_k} - \frac{1}{\hat{\mu}_{ij}} \right) & 1 \leq j < c_i, \\[10pt]
            0 & c_i \leq j \leq m,
        \end{cases}
    \end{align}
    where $c_i$ is the smallest integer index satisfying
    \begin{align}\label{def:c_i}
    \hat{\mu}_{ic_i} \left( \lambda_i + \sum_{k=1}^{c_i - 1} \frac{\mu_k}{\hat{\mu}_{ik}} \right) \leq \sum_{k=1}^{c_i - 1} \mu_k.
    \end{align}
\end{lem}
\begin{pf}
    Let us consider the Lagrangian function for player $i$, which is given by:
    \begin{align*}
        L_i(a, \alpha, \eta) = D_i(a) - \alpha \left( \sum_{j=1}^m a_{ij} - 1 \right) - \sum_{j=1}^m \eta_j a_{ij},
    \end{align*}
    where $\alpha$ and $\eta=(\eta_1,\ldots,\eta_m)$ are Lagrange multipliers corresponding to the constraints in \eqref{static-convex-programming}. Since \eqref{static-convex-programming} is a convex program, using KKT optimality conditions, we know that $a_{i}$ is the optimal solution (for a fixed $a_{-i}$) if and only if:
    \begin{align*}
        \frac{\partial L_i}{\partial a_{ij}} = 0 \quad  \forall j, \quad \frac{\partial L_i}{\partial \alpha} = 0,\\
        a_{ij} \eta_j = 0, \quad a_{ij} \geq 0, \quad \eta_j \geq 0 \quad \forall j.
    \end{align*}
    Using the first condition, we have:
    \begin{align*}
        \frac{\partial L_i}{\partial a_{ij}} = \frac{\lambda_i}{\hat{\mu}_{ij}} + \frac{\lambda_i^2 a_{ij}}{\mu_j} - \alpha - \eta_j = 0.
    \end{align*}
    If \(a_{ij} = 0\), we know that $\eta_j \geq 0$, which results in:
    \begin{align*}
        \alpha \leq \frac{\lambda_i}{\hat{\mu}_{ij}}.
    \end{align*}
    If \(a_{ij} \neq 0\), we have
    \begin{align*}
        \eta_j = 0 \implies \alpha = \lambda_i \left(\frac{1}{\hat{\mu}_{ij}}+ \frac{\lambda_i a_{ij}}{\mu_j} \right).
    \end{align*}
    Suppose we sort the servers according to their available processing rates such that
    \begin{align*}
        \hat{\mu}_{i1} \geq \hat{\mu}_{i2} \geq \dots \geq \hat{\mu}_{im}.
    \end{align*}
Now, a key observation is that a server with a higher available processing rate should naturally have a higher fraction of jobs assigned to it. Under the assumption that the servers are sorted by their available processing rates, the load fractions on the servers also should follow the same ordering, i.e., $a_{i1} \geq a_{i2} \geq \cdots \geq a_{im}$. Consequently, there may be cases where slower servers are assigned no jobs. This implies the existence of an index \(c_i\in [m]\) such that 
    \begin{align*}
        a_{ij} = 0 \quad \text{for } j = c_i, \ldots, m.
    \end{align*}
Therefore, suppose player $i$ only distributes its job among the servers $ 1 \leq j < c_i$. We have
    \begin{align*}
        &\alpha \sum_{k=1}^{c_i - 1} \mu_k = \lambda_i \sum_{k=1}^{c_i - 1} \frac{\mu_k}{\hat{\mu}_{ik}} + \lambda_i^2 \cr
        &\implies \alpha = \frac{\lambda_i (\lambda_i + \sum_{k=1}^{c_i - 1} \frac{\mu_k}{\hat{\mu}_{ik}})}{\sum_{k=1}^{c_i - 1} \mu_k}.
    \end{align*}
    Finally, for $1 \leq j < c_i$, we have
    \begin{align*}
        a_{ij} = \frac{\mu_j}{\lambda_i} \left( \frac{\lambda_i + \sum_{k=1}^{c_i - 1} \frac{\mu_k}{\hat{\mu}_{ik}}}{\sum_{k=1}^{c_i - 1} \mu_k} - \frac{1}{\hat{\mu}_{ij}} \right),
    \end{align*}
    where $c_i\in [m]$ is the smallest index satisfying
    \begin{align*}
        \frac{\lambda_i + \sum_{k=1}^{c_i - 1} \frac{\mu_k}{\hat{\mu}_{ik}}}{\sum_{k=1}^{c_i - 1} \mu_k} \leq \frac{1}{\hat{\mu}_{ic_i}}.
    \end{align*}$\hfill{\blacksquare}$
\end{pf}
\begin{rem}
  In practice, players may not have accurate estimates of the game parameters. To assess the robustness of best-response dynamics under observation noise, suppose the service rates are observed with noise: $\tilde{\mu}_k = \mu_k + \epsilon_k$, where $|\epsilon_k| \leq 1$. In this setting, one can show that the resulting best-response deviations are bounded in terms of the servers' service rates, provided the noise is small enough not to alter \(c_i\) as defined in Eq.~\eqref{def:c_i}.
\end{rem}
In Theorem~\ref{static-convergence-NE}, we established that the best-response dynamics in the static game converge to a pure NE. In fact, using the closed-form characterization of the best responses in Lemma~\ref{br-static}, we can prove a stronger result regarding the convergence time of the best-response dynamics to a pure NE in the static game. More specifically, we can show that under a mild assumption on the initial strategies, if each player updates their strategy exactly once using the best-response update rule \eqref{bestResponsePolicyStatic}, the game will reach a pure NE. This result is stated formally in the following theorem.

\begin{thm}\label{thm:stateic-n-iteration}
Suppose in the static load balancing game each player is initially distributing a non-zero portion of its job on each server, i.e., $a_{ij}^0 \neq 0, \ \forall i,j$. Then, the action profile obtained when players update their actions exactly once (in any order) using the best-response rule in Lemma~\ref{br-static} is a pure NE. In other words, the best-response dynamics converge to a pure NE in $n$ iterations.  
\end{thm}
\begin{pf}
Please see Appendix~\ref{appx:static-rate} for the proof.  
\end{pf}
We now proceed to bound the Price of Anarchy (PoA) of the static load balancing game, which is defined as the ratio between the worst-case social cost at a pure NE and the optimal social cost:
\begin{align*}
\text{PoA} = \frac{\max_{a \in \text{NE}} \sum_{i=1}^n D_i(a)}{\min_{a} \sum_{i=1}^n D_i(a)}.
\end{align*}
The PoA quantifies the inefficiency of Nash equilibria in games due to the selfish behavior of the players, compared to a globally optimal solution~\cite{Koutsoupias1999}.


\begin{thm}\label{thm:PoA}
The price of anarchy of the static load balancing game is bounded by 
    \[
    \text{PoA} \leq 1 + 2 \left(\max_{j\in[m]}\frac{s_j^0}{\mu_j} + \frac{\sum_{i=1}^n \lambda_i}{\mu_{\mathrm{min}}}\right)
    \left(\frac{\sum_{j=1}^m \mu_j}{\sum_{i=1}^n \lambda_i}\right).\]
\end{thm}
\begin{pf}
Please see Appendix~\ref{PoA-nonzero} for the proof.   
\end{pf}
\begin{rem}
    In fact, the bound obtained in Theorem~\ref{thm:PoA} can be further improved for special cases, such as when the initial loads are all zero, i.e., $s_j^0 = 0$ for all $j \in [m]$. In this case, as a direct corollary of the proof of Theorem~\ref{thm:PoA}, one can show that $\text{PoA} \leq 3$. We refer to Corollary~\ref{corr:PoA-special} in Appendix~\ref{PoA-nonzero} for a formal proof of this result.
\end{rem}

\section{Theoretical Results for the Dynamic Game}
\label{sec:dynamic}
In this section, we consider the dynamic load balancing game and present our main theoretical results. As we saw in Section \ref{sec:static}, the static load balancing game admits a pure NE, and iterative best-response dynamics converge to one such NE point. Therefore, a natural question arises: will following the best-response dynamics in the dynamic game also lead to a NE? The main result of this section is to show that this is indeed the case, and to establish a bound on the number of iterations required for the dynamics to converge to a fully balanced distribution of loads across the servers.

\begin{algorithm}[H]\caption{Best-Response Dynamics for the Dynamic Load Balancing Game}\label{alg:main} 
\smallskip
Arrival of a new job at time $t$ to player $i$ with length $\lambda_i$: 
\begin{itemize}
 \item Player $i$ observes state $s^t$ and computes available rates $\hat{\mu}^t_j = \frac{\mu_j}{s_j^t}, \, j\in[m]$.
 \item Sort servers so that $\hat{\mu}^t_1 \ge \hat{\mu}^t_2 \ge \cdots \ge \hat{\mu}^t_m$.
 \item Compute best response $a_i^t$ by solving $\min \{ D_i(a_i, s^t) : a_i \in A_i \}$, using Eqs.~\eqref{bestResponsePolicyStatic} and~\eqref{def:c_i}.
\item Schedule portions $\lambda_i a^t_{ij}$ of the job on server $j \in [m]$.
\item Proceed to next time step $t+1$.
\end{itemize}
\end{algorithm}
Algorithm~\ref{alg:main} describes the best-response dynamics for the dynamic load balancing game. In this algorithm, at each time $t$, the player receiving a job distributes it among the servers according to its best-response strategy relative to the current state. The following lemma establishes a key property of the update rule in Algorithm~\ref{alg:main}.
\begin{lem}\label{once-send-always-send}
    Suppose each player follows Algorithm~\ref{alg:main} in the dynamic load balancing game. Then, there exists a finite time $t'=\max_{j\in [m]} \lceil s^0_j/\mu_j\rceil$ after which any player must distribute its job among all the servers, i.e., $a^{t}_{ij}>0\  \forall i,j, t\ge t'$.
\end{lem}
\begin{pf}
Following Algorithm~\ref{alg:main}, suppose at an arbitrary time $t$ player \(i\) distributes its job to the set of servers indexed by \(\{1, \dots, c-1\}\), while sending nothing to the servers in \(\{c, \dots, m\}\), i.e.,
\begin{align}\label{eq:a-cases}
    a^t_{ij} = 
    \begin{cases} 
        \frac{\mu_j}{\lambda_i} \left( \frac{\lambda_i + \sum_{k=1}^{c - 1} s^t_k}{\sum_{k=1}^{c -1 }\mu_k} - \frac{s^t_j}{\mu_j} \right) & 1 \leq j < c, \\[10pt]
        0 & c \leq j \leq m,
    \end{cases}
\end{align}
where $c$ is the smallest index satisfying:
\begin{align}\label{c_samllest_index}
    \hat{\mu}^t_{c} \left( \lambda_i + \sum_{k=1}^{c -1} s_k \right) \leq \sum_{k=1}^{c - 1} \mu_k.
\end{align} 
At time \(t\), let us define
\begin{align}
\label{C_definition}
    C^t = \frac{\lambda_i + \sum_{k=1}^{c-1} s^t_k}{\sum_{k=1}^{c-1}\mu_k}, 
\end{align}
which in view of \eqref{eq:a-cases} implies that
\begin{align}\label{eq:C-a}
    \frac{\lambda_i a_{ij}^t + s_j^t}{\mu_j} = C^t \quad 1 \leq j < c.
\end{align}
Therefore, according to Eq.~\eqref{eq:state-change}, the states of the servers evolve as
\begin{align}
\label{s/mu_send}
    \frac{s_j^{t+1}}{\mu_j} = \frac{(\lambda_i a_{ij}^t + s_j^t - \mu_j)^+}{\mu_j} = (C^{t}-1)^+ \quad 1 \leq j < c,
\end{align}
and
\begin{align}
\label{s/mu_notsend}
    \frac{s_j^{t+1}}{\mu_j} = \frac{(s_j^t-\mu_j)^+}{\mu_j} \geq (C^{t}-1)^+ \quad c\leq  j \leq m,
\end{align}
where the last inequality is obtained using the update rule of Algorithm~\ref{alg:main} as for $c\leq j\leq m$, we have $\frac{s^t_j}{\mu_j}\ge C^t$.

Now, either $C^t \leq 1$, in which case by \eqref{s/mu_send} 
all the servers in \(\{1, \dots, c-1\}\) become empty at time $t+1$, or $C^t > 1$, in which case by \eqref{s/mu_send} and \eqref{s/mu_notsend}, the servers would still have some load remaining on them at time $t+1$. In either case, using \eqref{s/mu_send} and \eqref{s/mu_notsend}, we have
\begin{align}\label{eq:s-c-order}
    \frac{s_1^{t+1}}{\mu_1} = \dots = \frac{s_{c-1}^{t+1}}{\mu_{c-1}}=(C^{t}-1)^+\leq \frac{s_{c}^{t+1}}{\mu_{c}} \leq \dots \leq \frac{s_{m}^{t+1}}{\mu_{m}},  
\end{align}
or equivalently, 
\begin{align}
    \label{mu_sorted_t+1}
    \hat{\mu}^{t+1}_m \leq \cdots \leq \hat{\mu}^{t+1}_c \leq \hat{\mu}^{t+1}_{c-1} = \cdots = \hat{\mu}^{t+1}_{1}.
\end{align}
This shows that once a player distributes its job among a set of servers, those servers will continue to receive jobs in future iterations simply because the order of those servers in terms of available processing rates remains on the top of the list. More precisely, let us denote the set of servers that receive a job at time $t$ by $S^t=\{1,2,\ldots,c-1\}$, and suppose another player (say player $i'$) updates at time $t+1$ and wants to distribute its job among the servers. Then, according to Algorithm~\ref{alg:main}, player $i'$ would send its job to all the servers in $S^t$ because they all have the highest order in the list \eqref{mu_sorted_t+1}, and possibly more servers examined in the same order as \eqref{mu_sorted_t+1}. Therefore, the set of servers receiving a job at time $t+1$ is a superset of those that receive a job at time $t$, and they form a nested sequence of sets over time, i.e.,
\begin{align}\label{eq:nested}
S^t\subseteq S^{t+1}\subseteq S^{t+2}\subseteq \ldots.  
\end{align}
Next, we argue that the nested sequence \eqref{eq:nested} must grow until it eventually includes all the servers, i.e., $S^{t'}=[m]$ for some $t'$. In particular, using \eqref{eq:nested}, we must have $S^t=[m]\ \forall t\ge t'$, which means that all the servers must receive a job after time $t'$, completing the proof.

To show that the nested sequence \eqref{eq:nested} must grow, we first note that if a server $j$ becomes empty at some time $t$, i.e., $s_j^{t} = 0$, then $\hat{\mu}_j^{t} = \infty$. Thus, server $j$ must receive a job at the next time step $t+1$, and so by the nested property \eqref{eq:nested}, it must receive a job in all subsequent iterations. Since $c\notin S^{t}$, it means that in all iterations $1,2,\ldots,t$, server $c$ did not receive a new job and was only processing its initial load $s^0_c$ at rate $\mu_c$. Otherwise, server $c$ must have previously received a job and thus would be in the set $S^t$ by the nested property \eqref{eq:nested}. Therefore, the load of server $c$ at time $t$ would be $s_c^0-t\mu_c$. As a result, server $c$ does not belong to $S^t$ for at most $t\leq \lceil s^0_c/\mu_c\rceil$ iterations, after which its load becomes zero and according to the above discussion, it will receive a job in the next iteration. By repeating this argument for any choice of server $c$, one can see that after at most $t'=\max_{j\in [m]} \lceil s^0_j/\mu_j\rceil$ iterations, we must have $S^{t'}=[m]$. $\hfill{\blacksquare}$
\end{pf}
We are now ready to state the main result of this section.
\begin{thm}
\label{dynamic-convergence-NE}
If all the players follow Algorithm~\ref{alg:main}, after $$t''=O\left( \max_{j} \lceil \frac{s^0_j}{\mu_j}\rceil  \frac{\sum_{j}\mu_j+\sum_j s_j^0}{\sum_{j}\mu_j-\lambda_{\max}}\right)$$ iterations,\footnote{We refer to Appendix~\ref{appx:alternative-bound} for an alternative bound on the number of iterations $t''$, which may be better for a certain range of parameters.} the state of each server becomes zero (i.e., the system becomes load-balanced). In particular, the strategies of the players converge to the pure NE of the static load balancing game with zero initial loads.
\end{thm}
\vspace{-0.4cm}
\begin{pf}
    From Lemma~$\ref{once-send-always-send}$, we know that once a player begins sending jobs to all servers, all subsequent players will continue to do so for the remainder of the time. Consider an arbitrary time $t\ge t'$, and assume that at time $t$ player $i$ receives a job and plays its best-response according to Algorithm~\ref{alg:main}. Since all servers receive jobs at time $t\ge t'$, we have
    \begin{align*}
    a^t_{ij} = \frac{\mu_j}{\lambda_i} \left( \frac{\lambda_i + \sum_{k=1}^{m} s^t_k}{\sum_{k=1}^{m}\mu_k} - \frac{s^t_j}{\mu_j} \right),
    \end{align*}
    \begin{align*}
    (C^{t}-1)^+ = \frac{s_1^{t+1} + \dots + s_m^{t+1}}{\mu_1 + \dots + \mu_m}.
    \end{align*}
    Moreover, if player $i'$ schedules its job at time $t+1$, then
    \begin{align}\label{eq:C-recursive}
    C^{t+1}=\frac{\lambda_{i'} + \sum_{k=1}^{m} s^{t+1}_k}{\sum_{k=1}^{m}\mu_k} =\frac{\lambda_{i'}}{\sum_{k=1}^m\mu_k}+ (C^t - 1)^+ .
    \end{align}
There are two possible cases:
    \begin{itemize}
        \item If $C^t > 1$, then $\sum_{j=1}^{m} s_j^{t+1} \neq 0$. As in Eq.~\eqref{s/mu_send} where $c$ is replaced by $m$, we have
      \begin{align}\label{eq:s-drift}
            \frac{s_j^{t+1}}{\mu_j} = \frac{(\lambda_i a_{ij}^t + s_j^t - \mu_j)^+}{\mu_j} = (C^{t}-1)^+ \ \forall j\in [m]. 
            \end{align}
            Now because $C^t > 1$, we have $(C^t-1)^+>0$, which implies $(\lambda_i a_{ij}^t + s_j^t - \mu_j)^+ > 0$. Therefore, $(\lambda_i a_{ij}^t + s_j^t - \mu_j)^+\!=\! \lambda_i a_{ij}^t + s_j^t - \mu_j\ \forall j\in [m]$. Using this relation in \eqref{eq:s-drift} and summing over $j$ we get 
            \begin{align*}
        \sum_{j=1}^{m} s_j^{t+1} = \sum_{j=1}^{m} s_j^{t}-( \sum_{j=1}^m \mu_j-\lambda_i),
            \end{align*} 
            where we note that by the stability assumption, we have $0< \sum_{j=1}^m \mu_j-\lambda_i$.
                
        \item If $C^t \leq 1$, then $\sum_{j=1}^{m} s_j^{t+1} = 0$, and thus using \eqref{eq:C-recursive}, we have $C^{t+1} = \frac{\lambda_{i'}}{\sum_{k=1}^m \mu_k} < 1$. By repeating this argument inductively, one can see that $\sum_{j=1}^m s^{t+k}_j=0\ \forall k\ge 1$.   
    \end{itemize}
    Using the above cases, one can see that if $\sum_{j=1}^m s^{t+1}_j\neq 0$ for some $t\ge t'$, then the value of $\sum_{j=1}^m s^{t}_j$ decreases by at least  $\sum_{j=1}^m \mu_j-\lambda_{\max}>0$ in the next time step, where $\lambda_{\max}=\max_i \lambda_i$. Therefore, after a finite number of iterations 
    \begin{align}\nonumber
    t''&\mathrel{:=} t'+\frac{\sum_{j=1}^m s^{t'}_j}{\sum_{j=1}^m \mu_j-\lambda_{\max}}\cr
    &\leq t'+\frac{\sum_{j=1}^m s^0_j+t'\lambda_{\max}}{\sum_{j=1}^m \mu_j-\lambda_{\max}}\cr
    &=O\left( \max_{j} \lceil \frac{s^0_j}{\mu_j}\rceil  \frac{\sum_{j}\mu_j+\sum_j s_j^0}{\sum_{j}\mu_j-\lambda_{\max}}\right),  
    \end{align}
    we must have $\sum_{j=1}^m s^{t+1}_j=0\ \forall t\ge t''$. In particular, for any time $t\ge t''$, the best-response action for the updating player $i$ takes a simple form of
    \begin{align*}
        a^t_{ij} = \frac{\mu_j}{\mu_1 + \dots + \mu_m}\ \ \ j\in [m], \forall t\ge t'',
    \end{align*}
    that is, each server receives a fraction of the job proportional to its service rate and the load on each server would become $\frac{\lambda_i}{(\mu_1 + \dots + \mu_m)}$. 

    Finally, we note that since $\sum_{j=1}^{m} s^t_j = 0$ for $t \geq t''$, the dynamic load balancing game reduces to the static load balancing game with zero initial loads, i.e., $s^0_j=0 \ \forall j\in [m]$. Since by Theorem~\ref{static-convergence-NE}, the best-response strategy in the static game converges to a pure NE, we obtain that $\lim_{t\to \infty} a^t_{ij} = \frac{\mu_j}{\mu_1 + \dots + \mu_m}$ for any $i\in [n]$ and $j \in [m]$, which is precisely the pure NE of the static game with zero initial loads.$\hfill{\blacksquare}$
\end{pf}
\begin{rem}
Note that the bound on $t''$ given in Theorem~\ref{dynamic-convergence-NE} is a worst-case estimate. In practice, for many parameter settings, the system may converge significantly faster than the theoretical bound suggests (see Table~\ref{tab:wait_times}). Appendix~\ref{appx:alternative-bound} provides an alternative bound that can be tighter for certain parameter ranges.
\end{rem}

\begin{rem}
\noindent
Theorem~\ref{dynamic-convergence-NE} shows that under the best-response policies (Algorithm~\ref{alg:main}), server loads vanish in polynomial time, and each player’s overall cost converges to the unique minimizer of their zero-state instantaneous cost (Eq.~\eqref{costFunctionSingaleAgent}). Since the overall cost is defined as the time-average of instantaneous costs (Eq.~\eqref{eq:cumulative-cost}), the cost for player~$i$ converges to $\lambda_i^2/(2\sum_{j=1}^m \mu_j)$. Thus, no player has an incentive to deviate, and the best-response policies form a NE, yielding a fully load-balanced system.
\end{rem}

\section{Numerical Experiments}\label{sec:numerical}

\begin{figure}[htbp]
    \centering
    \begin{subfigure}[b]{0.4\textwidth}
        \centering
        \includegraphics[width=\textwidth]{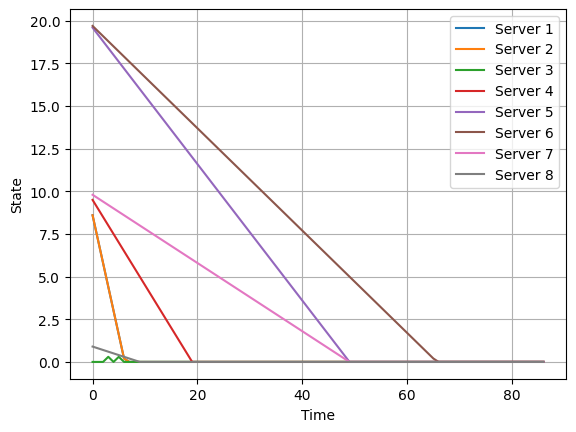}
        \caption{One player distributes a job at a time.}
        \label{fig:one_sending_setting1}
    \end{subfigure}
    \hfill
    \begin{subfigure}[b]{0.4\textwidth}
        \centering
        \includegraphics[width=\textwidth]{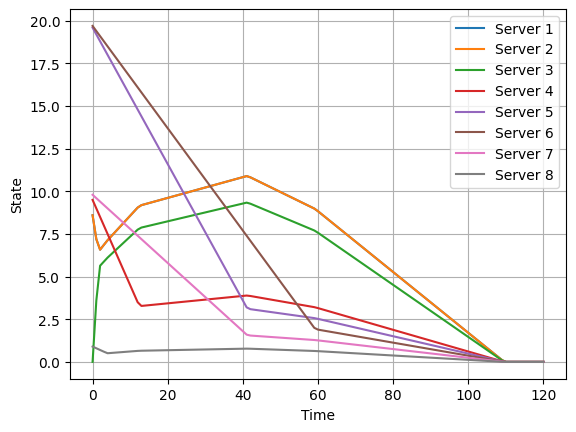}
        \caption{All players distribute jobs simultaneously.}
        \label{fig:all_sending_setting1}
    \end{subfigure}
    \hfill
    \begin{subfigure}[b]{0.4\textwidth}
        \centering
        \includegraphics[width=\textwidth]{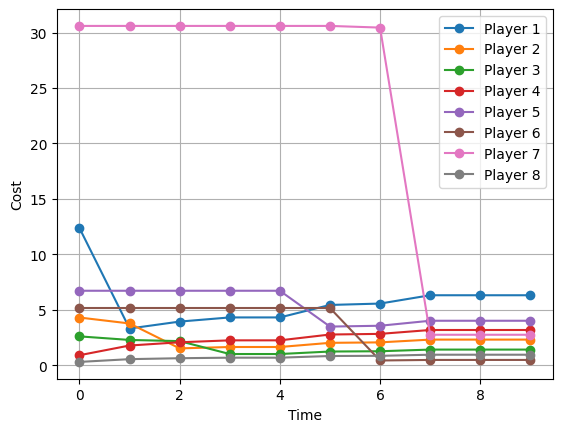}
        \caption{Sequentially playing best-response dynamics.}
        \label{fig:static_setting1}
    \end{subfigure}
    \caption{Comparison of different job distribution strategies for Setting 1, with parameters: $\mu = [1.4, 1.4, 1.2, 0.5, 0.4, 0.3, 0.2, 0.1], \lambda = [1.5, 0.5, 0.3, 0.7, \\0.9, 0.1, 0.6, 0.2], s^0 = [10, 10, 1, 10, 20, 20, 10, 1]$.}
    \label{fig:setting1}
\end{figure}

\begin{figure}[htbp]
    \centering
    \begin{subfigure}[b]{0.4\textwidth} 
        \centering
        \includegraphics[width=\textwidth]{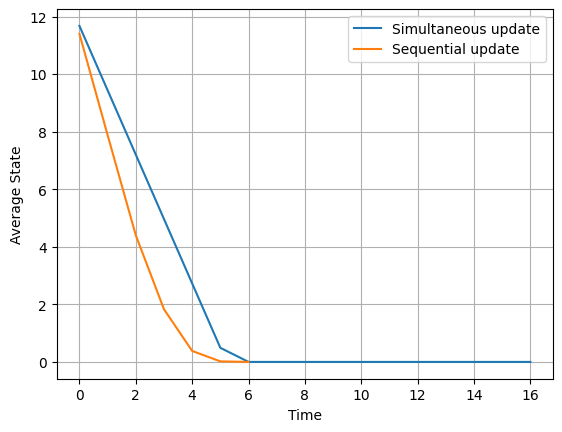} 
        \caption{Average server load under sequential and simultaneous updates of the dynamic game.}
        \label{fig:500-200-dynamic}
    \end{subfigure}
    \hfill 

    \vspace{0.1cm}
    \begin{subfigure}[b]{0.4\textwidth}
        \centering
        \includegraphics[width=\textwidth]{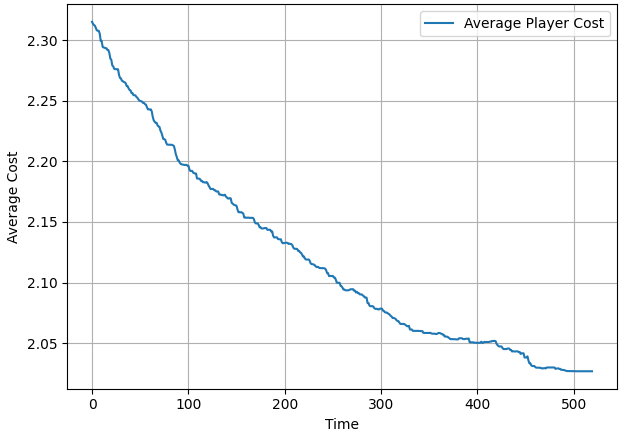}
        \caption{Average player cost in the static game.}
        \label{fig:500-200-static}
    \end{subfigure}
    \hfill
    \caption{Parameters for Setting 2: 500 players, 200 servers. Service rates $\sim \mathcal{U}[3,4]$, job lengths $\sim \mathcal{U}[2,3]$. Initial server states are sampled uniformly from $[10,20]$.}
    \label{fig:500p-200s}
\end{figure}

In this section, we present numerical results to evaluate the performance of the best-response dynamics under various settings of the static game, dynamic game, sequential update rule, and simultaneous update rule. In our simulations, we consider different choice of parameters for the initial state, players' job lengths, and servers' service rates, where the parameters for each setting are provided in the captions of the corresponding figures. For instance, in Fig.~\ref{fig:setting1}, we illustrate Setting 1, which involves eight servers with initial loads $[10, 10, 1, 10, 20, 20, 10, 1]$ and service rates $[1.4, 1.4, 1.2, 0.5, 0.4, 0.3, 0.2, 0.1]$, and a set of eight players with job lengths $[1.5, 0.5, 0.3, 0.7,
0.9, 0.1, 0.6, 0.2]$.

First, we simulate the performance of Algorithm~\ref{alg:main} for the dynamic load balancing game, where at each time step, a single player is randomly selected to distribute its job across the servers. The results of this experiment is illustrated in Fig.~\ref{fig:one_sending_setting1}. As can be seen, after sufficient time has elapsed, all states converge to zero. Once the load on all servers becomes zero, it remains at zero, indicating a stable NE. As illustrated in this figure, the convergence of the dynamics to the zero state appears to be fast and scales polynomially with respect to the game parameters, which also justifies the theoretical performance guarantee provided by Theorem~\ref{dynamic-convergence-NE}. 

Next, we evaluate an alternative version of Algorithm~\ref{alg:main} for the dynamic load balancing game, in which, at each time step, all players simultaneously distribute their jobs. In this case, each player adopts the best-response strategy, considering both the current state of the system and the strategies of other players. The result of this experiment is depicted in Fig.~\ref{fig:all_sending_setting1}. Interestingly, the results show that the simultaneous update of the players does not hurt the overall performance of the system and the state of the game eventually converges to zero, albeit at a slower convergence rate. 
 
Finally, in Fig.~\ref{fig:static_setting1}, we evaluate the performance of the sequential best-response dynamics in the static load balancing game, where $a^0_{ij} = \frac{1}{m}$ for all $i \in [n], j \in [m]$. As can be seen, the action profiles converge to a pure Nash equilibrium (NE) after exactly $n$ iterations (one update by each player), which confirms the theoretical convergence rate bound provided in Theorem~\ref{thm:stateic-n-iteration}.

To demonstrate the effectiveness of our framework in real-world scenarios, we evaluate an additional setting with significantly larger numbers of players and servers. Specifically, Setting 2 includes 500 players and 200 servers. In this case, each player's job lengths are drawn independently from the uniform distribution on $[2, 3]$, while the service rates of the servers are drawn uniformly from $[3, 4]$. Additionally, the initial state of each server is initialized by sampling uniformly from the interval $[10, 20]$. For this setting, we plot the average load on servers over time and compare the results for both sequential and simultaneous updates in the dynamic game. The results are depicted in Fig.~\ref{fig:500-200-dynamic}. Furthermore, we plot the average player cost over time for the sequential update in the static load balancing game, which is shown in Fig.~\ref{fig:500-200-static}.

\begin{figure}[htbp]
    \centering
    \begin{subfigure}[b]{0.4\textwidth} 
        \centering
        \includegraphics[width=\textwidth]{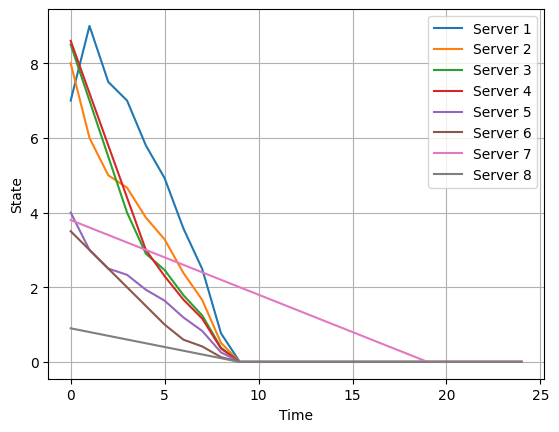} 
        \caption{$\mu = [3, 2, 1.5, 1.4, 1, 0.5, 0.2, 0.1]$, $\lambda = [2.5, 3, 4, 6, 5, 5, 5, 5]$, $s^0 = [10, 10, 10, 10, 5, 4, 4, 1]$}
        \label{fig:one_sending_setting4}
    \end{subfigure}
    \hfill 

    \vspace{0.1cm}
    \begin{subfigure}[b]{0.4\textwidth}
        \centering
        \includegraphics[width=\textwidth]{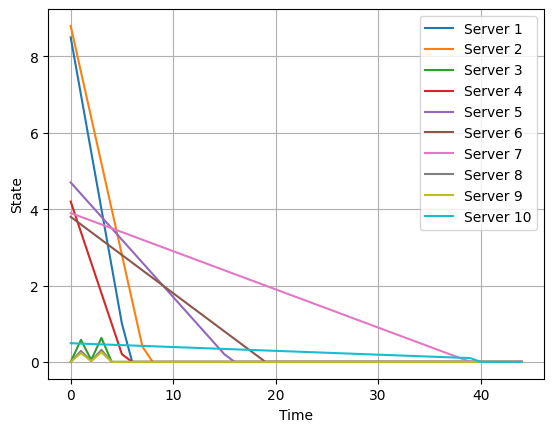}
        \caption{$\mu = [1.5, 1.2, 1.0, 0.8, 0.3, 0.2, 0.1, 0.5\\, 0.4, 0.01]$, $\lambda = [4, 0.8, 0.9, 0.7, 0.6, 0.5, 0.4, 0.3, 3]$, $s^0 = [10, 10, 1, 5, 5, 4, 4, 0.1, 0.1, 0.5]$}
        \label{fig:one_sending_setting5}
    \end{subfigure}
    \hfill
    \caption{One player distributes its job at a time.}
    \label{parameter4&5}
\end{figure}

It is also worth noting that certain parameter settings are feasible under the sequential update of the dynamic game but not under the simultaneous update. For instance, consider a system with service rates $[3, 2, 1.5, 1.4, 1, 0.5, 0.2, 0.1]$ and job lengths $[2.5, 3, 4, 6, 5, 5, 5, 5]$. This setting is feasible for sequential updates because, for each player $i$, we have $\lambda_i < \sum_{j=1}^8 \mu_j = 9.7$. However, it becomes infeasible under simultaneous updates since the total arrival rate exceeds the total service rate: $\sum_{i=1}^8 \lambda_i = 35.5 > \sum_{j=1}^8 \mu_j = 9.7$. Figure~\ref{parameter4&5} illustrates the performance of Algorithm~\ref{alg:main} for two such cases.

\begin{figure}[htbp]
    \centering
    \includegraphics[width=0.4\textwidth]{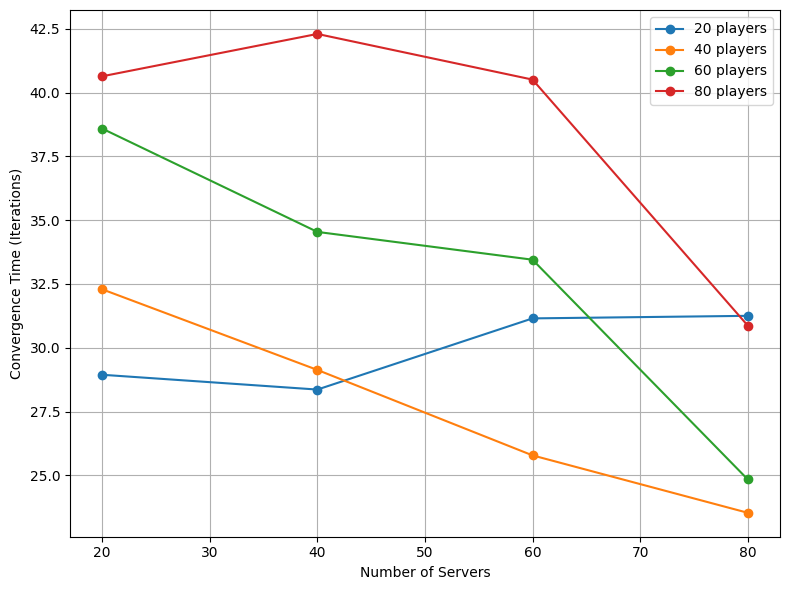}
    \caption{Convergence time for varying numbers of players and servers.}
    \label{fig:convergence-varying-n-m}
\end{figure}

Next, we further extend our numerical results by systematically varying the number of players ($n$) and servers ($m$), and analyzing how the convergence time of Algorithm~\ref{alg:main} scales with system size. In this experiment, both $n$ and $m$ are varied over the set $\{20, 40, 60, 80\}$. The service rates $\mu_j$ are sampled independently and uniformly from the interval $[1, 2]$, while job lengths are drawn uniformly from $[0, 1]$. The initial queue states are also initialized by sampling uniformly from $[10, 20]$. The results of this experiment are shown in Fig.~\ref{fig:convergence-varying-n-m}.

Finally in Table \ref{tab:wait_times} we compare the iteration counts of sequential and simultaneous update strategies across various settings in the dynamic game. Settings 1 and 2 are defined in Figs. \ref{fig:setting1} and \ref{fig:500p-200s} respectively. Settings 3-7 are defined as follows:
\begin{itemize}
    \item Setting 3: 100 players, 50 servers. Service rates $\sim \mathcal{U}[1,2]$, job lengths $\sim \mathcal{U}[0,1]$, initial states $\sim \mathcal{U}[10,20]$.
    \item Setting 4: 200 players, 100 servers. Service rates $\sim \mathcal{U}[2,3]$, job lengths $\sim \mathcal{U}[1,2]$, initial states $\sim \mathcal{U}[10,20]$.
    \item Setting 5: $\mu = [0.9, 0.8, 0.4, 0.01]$, $\lambda = [0.5, 0.5, 0.3, 0.7]$, $s^0 = [10, 10, 1, 0.5]$.
    \item Setting 6: $\mu = [1.4, 1.2, 1, 0.5]$, $\lambda = [0.5, 0.5, 0.3, 0.7, 0.9,\\ 0.1, 0.6, 0.2]$, $s^0 = [10, 10, 1, 10]$.
    \item Setting 7: $\mu = [2, 2]$, $\lambda = [0.5, 0.5, 0.3, 0.7, 0.9, 0.1, 0.6,\\ 0.2]$, $s^0 = [10, 50]$.
\end{itemize}
\begin{table}[h]
\centering
\small  
\scalebox{0.9}{
\begin{tabular}{|c|c|c|}
\hline
\textbf{Setting} & \textbf{Iter. count (simult.)} & \textbf{Iter. count (seq.)} \\
\hline
1 & 110 & 66\\
\hline
2 & 6 & 6\\
\hline
3 & 32 &  19 \\
\hline
4 & 9 &  8 \\
\hline
5 &  176 &  45 \\
\hline
6 & 90  & 19  \\
\hline
7 & 280  & 24  \\
\hline
\end{tabular}}
\caption{Iteration counts for simultaneous vs. sequential update strategies under various settings.}
\label{tab:wait_times}
\end{table}
\section{Conclusions}
\label{sec:conclusion}
In this work, we develop a game-theoretic framework for load balancing in static and dynamic settings, motivated by practical applications in job scheduling and cyber-physical systems. We demonstrate that the static game formulation is a potential game, which guarantees the existence of a pure Nash equilibrium, with best-response dynamics converging in a finite number of iterations. We then bound the price of anarchy (PoA) of the static game in terms of game parameters. To account for the carryover effect, we extend our formulation to dynamic settings and show that players employing best-response strategies in each state can achieve a fully load-balanced system, with convergence to the Nash equilibrium of the static game occurring in polynomial time. Our numerical experiments validate the efficiency of the proposed algorithms. 

These findings provide a solid foundation for further exploration into more complex load-balancing strategies in distributed systems. One promising direction for future work is to explore scenarios where all players update their strategies and distribute jobs simultaneously in each round of the dynamic game. While our numerical experiments suggest that the dynamic game would still converge to a load-balanced static game, the convergence rate is notably slower. Another interesting direction is to extend our work to stochastic settings, where servers have stochastic processing rates and process the jobs assigned to them with certain probabilities. 

\bibliographystyle{unsrt}
\bibliography{autosam}           

@article{grosu2005noncooperative,
  title={Noncooperative load balancing in distributed systems},
  author={Grosu, Daniel and Chronopoulos, Anthony T},
  journal={Journal of Parallel and Distributed Computing},
  volume={65},
  number={9},
  pages={1022--1034},
  year={2005},
  publisher={Elsevier}
}

@inproceedings{gaitonde2020stability,
  title={Stability and learning in strategic queuing systems},
  author={Gaitonde, Jason and Tardos, {\'E}va},
  booktitle={Proceedings of the 21st ACM Conference on Economics and Computation},
  pages={319--347},
  year={2020}
}

@inproceedings{gaitonde2021virtues,
  title={Virtues of patience in strategic queuing systems},
  author={Gaitonde, Jason and Tardos, Eva},
  booktitle={Proceedings of the 22nd ACM Conference on Economics and Computation},
  pages={520--540},
  year={2021}
}

@article{monderer1996potential,
  title={Potential games},
  author={Monderer, Dov and Shapley, Lloyd S},
  journal={Games and Economic Behavior},
  volume={14},
  number={1},
  pages={124--143},
  year={1996},
}

@article{schaerf1994adaptive,
  title={Adaptive load balancing: A study in multi-agent learning},
  author={Schaerf, Andrea and Shoham, Yoav and Tennenholtz, Moshe},
  journal={Journal of Artificial Intelligence Research},
  volume={2},
  pages={475--500},
  year={1994}
}

@article{krishnasamy2016regret,
  title={Regret of queueing bandits},
  author={Krishnasamy, Subhashini and Sen, Rajat and Johari, Ramesh and Shakkottai, Sanjay},
  journal={Advances in Neural Information Processing Systems},
  volume={29},
  year={2016}
}

@article{milchtaich1996congestion,
  title={Congestion games with player-specific payoff functions},
  author={Milchtaich, Igal},
  journal={Games and Economic Behavior},
  volume={13},
  number={1},
  pages={111--124},
  year={1996},
}

@article{holzman1997strong,
  title={Strong equilibrium in congestion games},
  author={Holzman, Ron and Law-Yone, Nissan},
  journal={Games and Economic Behavior},
  volume={21},
  number={1-2},
  pages={85--101},
  year={1997},
}

@inproceedings{grosu2002load,
  title={Load balancing in distributed systems: An approach using cooperative games},
  author={Grosu, Daniel and Chronopoulos, Anthony T and Leung, Ming-Ying},
  booktitle={In Proceedings of the 16th  IEEE International Parallel and Distributed Processing Symposium},
  pages={10--pp},
  year={2002},
}

@book{kameda2012optimal,
  title={Optimal load balancing in distributed computer systems},
  author={Kameda, Hisao and Li, Jie and Kim, Chonggun and Zhang, Yongbing},
  year={2012},
  publisher={Springer Science \& Business Media}
}

@inproceedings{roughgarden2001stackelberg,
  title={Stackelberg scheduling strategies},
  author={Roughgarden, Tim},
  booktitle={Proceedings of the 33rd Annual ACM Symposium on Theory of Computing},
  pages={104--113},
  year={2001}
}

@article{nash1950bargaining,
  title={The bargaining problem},
  author={Nash Jr, John F},
  journal={Econometrica: Journal of the Econometric Society},
  pages={155--162},
  year={1950},
}

@inproceedings{economides1991multi,
  title={Multi-Objective Routing in Integrated Services Networks: A Game Theory Approach.},
  author={Economides, Anastasios A and Silvester, John A and others},
  booktitle={Infocom},
  volume={91},
  pages={1220--1227},
  year={1991}
}

@inproceedings{economides1990game,
  title={A game theory approach to cooperative and non-cooperative routing problems},
  author={Economides, Anastasios A and Silvester, John A},
  booktitle={SBT/IEEE International Symposium on Telecommunications},
  pages={597--601},
  year={1990},
}

@article{orda1993competitive,
  title={Competitive routing in multiuser communication networks},
  author={Orda, Ariel and Rom, Raphael and Shimkin, Nahum},
  journal={IEEE/ACM Transactions on Networking},
  volume={1},
  number={5},
  pages={510--521},
  year={1993},
}

@article{altman2001routing,
  title={Routing into two parallel links: Game-theoretic distributed algorithms},
  author={Altman, Eitan and Basar, Tamer and Jim{\'e}nez, Tania and Shimkin, Nahum},
  journal={Journal of Parallel and Distributed Computing},
  volume={61},
  number={9},
  pages={1367--1381},
  year={2001},
}

@article{korilis1997capacity,
  title={Capacity allocation under noncooperative routing},
  author={Korilis, Yannis A and Lazar, Aurel A and Orda, Ariel},
  journal={IEEE Transactions on Automatic Control},
  volume={42},
  number={3},
  pages={309--325},
  year={1997},
}

@inproceedings{koutsoupias1999worst,
  title={Worst-case equilibria},
  author={Koutsoupias, Elias and Papadimitriou, Christos},
  booktitle={Annual symposium on theoretical aspects of computer science},
  pages={404--413},
  year={1999},
}

@inproceedings{mavronicolas2001price,
  title={The price of selfish routing},
  author={Mavronicolas, Marios and Spirakis, Paul},
  booktitle={Proceedings of the thirty-third annual ACM Symposium on Theory of Computing},
  pages={510--519},
  year={2001}
}

@article{roughgarden2002bad,
  title={How bad is selfish routing?},
  author={Roughgarden, Tim and Tardos, {\'E}va},
  journal={Journal of the ACM (JACM)},
  volume={49},
  number={2},
  pages={236--259},
  year={2002},
}

@article{gupta2007analysis,
  title={Analysis of join-the-shortest-queue routing for web server farms},
  author={Gupta, Varun and Balter, Mor Harchol and Sigman, Karl and Whitt, Ward},
  journal={Performance Evaluation},
  volume={64},
  number={9-12},
  pages={1062--1081},
  year={2007},
}

@article{eschenfeldt2018join,
  title={Join the shortest queue with many servers. The heavy-traffic asymptotics},
  author={Eschenfeldt, Patrick and Gamarnik, David},
  journal={Mathematics of Operations Research},
  volume={43},
  number={3},
  pages={867--886},
  year={2018},
}

@article{foley2001join,
  title={Join the shortest queue: stability and exact asymptotics},
  author={Foley, Robert D and McDonald, David R},
  journal={Annals of Applied Probability},
  pages={569--607},
  year={2001},
}

@article{lu2011join,
  title={Join-Idle-Queue: A novel load balancing algorithm for dynamically scalable web services},
  author={Lu, Yi and Xie, Qiaomin and Kliot, Gabriel and Geller, Alan and Larus, James R and Greenberg, Albert},
  journal={Performance Evaluation},
  volume={68},
  number={11},
  pages={1056--1071},
  year={2011},
}

@inproceedings{mitzenmacher2016analyzing,
  title={Analyzing distributed join-idle-queue: A fluid limit approach},
  author={Mitzenmacher, Michael},
  booktitle={2016 54th Annual Allerton Conference on Communication, Control, and Computing (Allerton)},
  pages={312--318},
  year={2016},
}

@article{mitzenmacher2001power,
  title={The power of two choices in randomized load balancing},
  author={Mitzenmacher, Michael},
  journal={IEEE Transactions on Parallel and Distributed Systems},
  volume={12},
  number={10},
  pages={1094--1104},
  year={2001},
}

@article{xie2015power,
  title={Power of d choices for large-scale bin packing: A loss model},
  author={Xie, Qiaomin and Dong, Xiaobo and Lu, Yi and Srikant, Rayadurgam},
  journal={ACM SIGMETRICS Performance Evaluation Review},
  volume={43},
  number={1},
  pages={321--334},
  year={2015},
}

@article{gardner2017redundancy,
  title={Redundancy-d: The power of d choices for redundancy},
  author={Gardner, Kristen and Harchol-Balter, Mor and Scheller-Wolf, Alan and Velednitsky, Mark and Zbarsky, Samuel},
  journal={Operations Research},
  volume={65},
  number={4},
  pages={1078--1094},
  year={2017},
}

@inproceedings{kogias2019r2p2,
  title={$\{$R2P2$\}$: Making $\{$RPCs$\}$ first-class datacenter citizens},
  author={Kogias, Marios and Prekas, George and Ghosn, Adrien and Fietz, Jonas and Bugnion, Edouard},
  booktitle={2019 USENIX Annual Technical Conference (USENIX ATC 19)},
  pages={863--880},
  year={2019}
}

@inproceedings{kogias2020hovercraft,
  title={HovercRaft: Achieving scalability and fault-tolerance for microsecond-scale datacenter services},
  author={Kogias, Marios and Bugnion, Edouard},
  booktitle={Proceedings of the Fifteenth European Conference on Computer Systems},
  pages={1--17},
  year={2020}
}

@article{gardner2019smart,
  title={Smart dispatching in heterogeneous systems},
  author={Gardner, Kristen and Stephens, Cole},
  journal={ACM SIGMETRICS Performance Evaluation Review},
  volume={47},
  number={2},
  pages={12--14},
  year={2019},
}

@article{winston1977optimality,
  title={Optimality of the shortest line discipline},
  author={Winston, Wayne},
  journal={Journal of Applied Probability},
  volume={14},
  number={1},
  pages={181--189},
  year={1977},
}

@article{jennings2015resource,
  title={Resource management in clouds: Survey and research challenges},
  author={Jennings, Brendan and Stadler, Rolf},
  journal={Journal of Network and Systems Management},
  volume={23},
  number={3},
  pages={567--619},
  year={2015},
}

@inproceedings{ousterhout2013sparrow,
  title={Sparrow: distributed, low latency scheduling},
  author={Ousterhout, Kay and Wendell, Patrick and Zaharia, Matei and Stoica, Ion},
  booktitle={Proceedings of the twenty-fourth ACM Symposium on Operating Systems Principles},
  pages={69--84},
  year={2013}
}

@inproceedings{roy2015chaos,
  title={Chaos: Scale-out graph processing from secondary storage},
  author={Roy, Amitabha and Bindschaedler, Laurent and Malicevic, Jasmina and Zwaenepoel, Willy},
  booktitle={Proceedings of the 25th Symposium on Operating Systems Principles},
  pages={410--424},
  year={2015}
}

@inproceedings{nasir2015power,
  title={The power of both choices: Practical load balancing for distributed stream processing engines},
  author={Nasir, Muhammad Anis Uddin and Morales, Gianmarco De Francisci and Garcia-Soriano, David and Kourtellis, Nicolas and Serafini, Marco},
  booktitle={2015 IEEE 31st International Conference on Data Engineering},
  pages={137--148},
  year={2015},
}

@inproceedings{zhu2020racksched,
  title={$\{$RackSched$\}$: A $\{$Microsecond-Scale$\}$ Scheduler for $\{$Rack-Scale$\}$ Computers},
  author={Zhu, Hang and Kaffes, Kostis and Chen, Zixu and Liu, Zhenming and Kozyrakis, Christos and Stoica, Ion and Jin, Xin},
  booktitle={14th USENIX Symposium on Operating Systems Design and Implementation (OSDI 20)},
  pages={1225--1240},
  year={2020},
}

@article{zhou2018degree,
  title={Degree of queue imbalance: Overcoming the limitation of heavy-traffic delay optimality in load balancing systems},
  author={Zhou, Xingyu and Wu, Fei and Tan, Jian and Srinivasan, Kannan and Shroff, Ness},
  journal={Proceedings of the ACM on Measurement and Analysis of Computing Systems},
  volume={2},
  number={1},
  pages={1--41},
  year={2018}, 
}

@article{hellemans2018power,
  title={On the power-of-d-choices with least loaded server selection},
  author={Hellemans, Tim and Van Houdt, Benny},
  journal={Proceedings of the ACM on Measurement and Analysis of Computing Systems},
  volume={2},
  number={2},
  pages={1--22},
  year={2018},
}

@article{horvath2019mean,
  title={Mean field analysis of join-below-threshold load balancing for resource sharing servers},
  author={Horv{\'a}th, Ill{\'e}s Antal and Scully, Ziv and Van Houdt, Benny},
  journal={Proceedings of the ACM on Measurement and Analysis of Computing Systems},
  volume={3},
  number={3},
  pages={1--21},
  year={2019},
}

@article{lin1992dynamic,
  title={A dynamic load-balancing policy with a central job dispatcher (LBC)},
  author={Lin, Hwa-Chun and Raghavendra, Cauligi S},
  journal={IEEE Transactions on Software Engineering},
  volume={18},
  number={2},
  pages={148},
  year={1992},
}

@article{etesami2020smart,
  title={Smart routing of electric vehicles for load balancing in smart grids},
  author={Etesami, S Rasoul and Saad, Walid and Mandayam, Narayan B and Poor, H Vincent},
  journal={Automatica},
  volume={120},
  pages={109148},
  year={2020},
  publisher={Elsevier}
}

@book{kleinrock1975queueing,
  title     = {Queueing Systems, Volume I: Theory},
  author    = {Kleinrock, Leonard},
  year      = {1975},
  publisher = {Wiley-Interscience}
}

@article{cui2022learning,
  title={Learning in congestion games with bandit feedback},
  author={Cui, Qiwen and Xiong, Zhihan and Fazel, Maryam and Du, Simon S},
  journal={Advances in Neural Information Processing Systems},
  volume={35},
  pages={11009--11022},
  year={2022}
}

@article{krichene2015online,
  title={Online learning of nash equilibria in congestion games},
  author={Krichene, Walid and Drigh{\`e}s, Benjamin and Bayen, Alexandre M},
  journal={SIAM Journal on Control and Optimization},
  volume={53},
  number={2},
  pages={1056--1081},
  year={2015},
  publisher={SIAM}
}

@inproceedings{tennenholtz2009learning,
  title={Learning equilibria in repeated congestion games},
  author={Tennenholtz, Moshe and Zohar, Aviv},
  booktitle={Proceedings of The 8th International Conference on Autonomous Agents and Multiagent Systems-Volume 1},
  pages={233--240},
  year={2009}
}

@article{rosenthal1973class,
  title={A class of games possessing pure-strategy Nash equilibria},
  author={Rosenthal, Robert W},
  journal={International Journal of Game Theory},
  volume={2},
  number={1},
  pages={65--67},
  year={1973},
  publisher={Physica-Verlag Heidelberg}
}

@article{berenbrink2007distributed,
  title={Distributed selfish load balancing},
  author={Berenbrink, Petra and Friedetzky, Tom and Goldberg, Leslie Ann and Goldberg, Paul W and Hu, Zengjian and Martin, Russell},
  journal={SIAM Journal on Computing},
  volume={37},
  number={4},
  pages={1163--1181},
  year={2007},
  publisher={SIAM}
}

@incollection{kameda1997comparison,
  title={A comparison of static and dynamic load balancing},
  author={Kameda, Hisao and Li, Jie and Kim, Chonggun and Zhang, Yongbing},
  booktitle={Optimal Load Balancing in Distributed Computer Systems},
  pages={225--240},
  year={1997},
  publisher={Springer}
}

@article{abel2025learning,
  title={Learning in Strategic Queuing Systems with Small Buffers},
  author={Abel, Ariana and Kolumbus, Yoav and Duque, Jeronimo Martin and Foster, Cristian Palma and Tardos, Eva},
  journal={arXiv preprint arXiv:2502.08898},
  year={2025}
}

@inproceedings{Koutsoupias1999,
  author    = {Elias Koutsoupias and Christos Papadimitriou},
  title     = {Worst-case equilibria},
  booktitle = {Proceedings of the 16th Annual Symposium on Theoretical Aspects of Computer Science (STACS)},
  pages     = {404--413},
  year      = {1999},
  publisher = {Springer},
  doi       = {10.1007/3-540-49116-3_38}
}

@article{nash1950equilibrium,
  title={Equilibrium points in n-person games},
  author={Nash Jr, John F},
  journal={Proceedings of the national academy of sciences},
  volume={36},
  number={1},
  pages={48--49},
  year={1950},
  publisher={national academy of sciences}
}

\appendix
\section{Omitted Proofs} \label{appendix-proofs}
\subsection{Proof of Theorem~\ref{thm:stateic-n-iteration} }\label{appx:static-rate}
\begin{pf}
We prove the result by induction on sequential player updates. First, we show that the sets of servers receiving jobs form an increasing sequence as players best respond. Finally, we verify that the last player’s allocation equalizes the normalized loads, ensuring that no player can improve unilaterally.

Without loss of generality and by relabeling the players, let us assume that players update their strategies according to their index, i.e., the $i$th updating player is player $i$. Moreover, let us define $S^t_i$ to be the set of servers that receive a job from player $i$ given its action profile $a_i^t$ at time $t$. First, we show that if players sequentially update their actions according to their best responses, then $S^1_{1}\subseteq S^{2}_2 \cdots \subseteq S^{n}_n$.

Consider an arbitrary player $i$ who wants to update its action at time $t=i$. Since players $\{i+1,\ldots, n\}$ have not updated their actions yet, player $i$ plays its best response with respect to the action profile $\{a_1^1, \cdots, a_{i-1}^{i-1}, a_{i+1}^{0}, \cdots, a_{n}^{0}\}$. For each server $j$, define
\begin{align*}
\hat{\mu}^t_{ij} = \frac{\mu_j}{\sum_{k=1}^{i-1 }\lambda_k a^k_{kj} + \sum_{k=i+1}^{n}\lambda_k a^{0}_{kj}+ s_j^0},  
\end{align*}
which is the \emph{relative available processing rate} of server $j$ that is viewed by player $i$ at time $t = i$. Thus, player $i$ sorts servers in a way that $\hat{\mu}^t_{i1} \geq \hat{\mu}^t_{i2} \cdots \geq \hat{\mu}^t_{im}$. Let $c_i$ be the smallest integer index satisfying
    \begin{align}\label{C_less_than_mu}
        \frac{\lambda_i + \sum_{k=1}^{c_i - 1} \frac{\mu_k}{\hat{\mu}^t_{ik}}}{\sum_{k=1}^{c_i - 1}\mu_k} \leq \frac{1}{\hat{\mu}^t_{i,c_i}},
    \end{align}
and define $C_i^t \coloneqq \frac{\lambda_i + \sum_{k=1}^{c_i - 1} \frac{\mu_k}{\hat{\mu}^t_{ik}}}{\sum_{k=1}^{c_i - 1}\mu_k}$. By the best-response rule, player $i$ sends jobs to servers $\{1,\dots,c_i-1\}$. For these servers we have
    \[
    C_i^t = \frac{\lambda_i a_{ij}^t}{\mu_j} + \frac{1}{\hat{\mu}_{ij}^t} \quad 1 \leq j < c_i.
    \]
Now when player $i+1$ updates, the inverse available rates satisfy
    \[
    \frac{1}{\hat{\mu}^{t+1}_{i+1,j}} =
    \begin{cases}
    C_i^t - \frac{\lambda_{i+1}a^{0}_{i+1,j}}{\mu_j} & 1 \leq j < c_i, \\
    \frac{1}{\hat{\mu}^t_{ij}} - \frac{\lambda_{i+1}a^{0}_{i+1,j}}{\mu_j} & c_i \leq j \leq m.
    \end{cases}
    \]
There are two possible scenarios:
    \begin{itemize}
        \item $c_{i} - 1=m$: This means that player $i$ distributes its job across all servers and we now want to check if that is also the case for player $i+1$. Suppose server $k$ is the server with highest $\hat{\mu}^{t+1}_{i+1,k}$, so that we know player $i+1$ sends jobs to it. Now, we have to check whether player $i$ also sends a job to other servers. Take server $h \in [m]\setminus\{k\}$ as an example. Player $i+1$ sends a portion of its job to server $h$ only if
        \begin{align}\nonumber
            C_i^t <& \frac{\lambda_{i+1} a^{0}_{i+1, h}}{\mu_h} + \frac{\lambda_{i+1} + \frac{\mu_k}{\hat{\mu}^{t+1}_{i+1,k}}}{\mu_k} \cr 
            =& 
            \frac{\lambda_{i+1} a^{0}_{i+1, h}}{\mu_h} +\frac{\lambda_{i+1}}{\mu_k} + C_i^t - \frac{\lambda_{i+1} a^{0}_{i+1, k}}{\mu_k},
        \end{align}
         which holds because we assume that for all $j\in[m]$, $a_{i+1, j}^0 \neq 0$. More generally, for all $h \in [m]$ and $\mathcal{M}_{i+1}^{h} \subseteq [m]\setminus \{h\}$, we have
        \begin{align}\nonumber
            C_i^t <& \frac{\lambda_{i+1} a^{0}_{i+1, h}}{\mu_h} + \frac{\lambda_{i+1} + \sum_{j \in \mathcal{M}_{i+1}^{h}} \frac{\mu_j}{\hat{\mu}^{t+1}_{i+1, j}}}{\sum_{j \in \mathcal{M}_{i+1}^{h}} \mu_j} \cr 
            =& 
            \frac{\lambda_{i+1} a^{0}_{i+1, h}}{\mu_h} +\frac{\lambda_{i+1}}{\sum_{j \in \mathcal{M}_{i+1}^{h}} \mu_j} + C_i^t \cr-& \frac{\lambda_{i+1} \sum_{j \in \mathcal{M}_{i+1}^{h}} a^{0}_{i+1, j}}{\sum_{j \in \mathcal{M}_{i+1}^{h}} \mu_j},
        \end{align}
        which holds because for all $i \in [n], j \in [m]$, $a_{i+1,j}^{0}\neq 0$. Thus, we conclude that $c_{i+1}-1 = m$, which is equivalent to $S^{t+1}_{i+1} = S^t_i$.

        \item $c_{i} - 1 < m$: Suppose player $i+1$ assigns jobs to the set of servers $\mathcal{N}^{t+1}_{i+1} \subseteq \{c_i, \dots, m\}$. Note that $\mathcal{N}^{t+1}_{i+1}$ may also be empty. We now examine whether player $i+1$ assigns any jobs to the servers in $S^t_i$. Consider, for example, a server $k \in S^t_i = \{1, \dots, c_i - 1\}$. Player $i+1$ assigns jobs to server $k$ only if
        \begin{align}\nonumber
            C_i^t <& \frac{\lambda_{i+1} a^{0}_{i+1, k}}{\mu_k} + \frac{\lambda_{i+1} + \sum_{j \in \mathcal{N}^{t+1}_{i+1}} \frac{\mu_j}{\hat{\mu}^{t+1}_{i+1,j}}}{\sum_{j \in \mathcal{N}^{t+1}_{i+1}} \mu_j} \cr 
            =& 
            \frac{\lambda_{i+1} a^{0}_{i+1, k}}{\mu_k} + \frac{\lambda_{i+1} - \lambda_{i+1} \sum_{j \in \mathcal{N}^{t+1}_{i+1}} a^{0}_{i+1, j}}{\sum_{j \in \mathcal{N}^{t+1}_{i+1}} \mu_j} \cr+& \frac{\sum_{j \in \mathcal{N}^{t+1}_{i+1}} \frac{\mu_j}{\hat{\mu}^t_{ij}}}{\sum_{j \in \mathcal{N}^{t+1}_{i+1}} \mu_j}.
        \end{align}
        According to \eqref{C_less_than_mu}, we already know that 
        \begin{align}\label{C_less_sum_mu}
            C_i^t \leq \frac{\sum_{j \in \mathcal{N}^{t+1}_{i+1}} \frac{\mu_j}{\hat{\mu}^{t}_{i,j}}}{\sum_{j \in \mathcal{N}^{t+1}_{i+1}} \mu_j}.
        \end{align}
        Since $a^{0}_{i+1,j} \neq 0$ for all $j \in [m]$, we conclude that player $i+1$ sends a fraction of its job to server $k$ at time $t+1$. To check whether player $i+1$ also assigns jobs to the remaining servers in $S_i^t$, consider an arbitrary server $h \in S_i^t \setminus \{k\}$. Player $i+1$ assigns jobs to server $h$ only if
        \begin{align}\nonumber
            C_i^t <& \frac{\lambda_{i+1} a^{0}_{i+1, h}}{\mu_h} + \frac{\lambda_{i+1} + \frac{\mu_k}{\hat{\mu}^{t+1}_{i+1,k}} + \sum_{j \in \mathcal{N}^{t+1}_{i+1}} \frac{\mu_j}{\hat{\mu}^{t+1}_{i+1,j}}}{\mu_k + \sum_{j \in \mathcal{N}^{t+1}_{i+1}} \mu_j} \cr
            =& \frac{\lambda_{i+1} a^{0}_{i+1, h}}{\mu_h} + \frac{\lambda_{i+1} + C_i^t \mu_k -\lambda_{i+1} a_{i+1, k}^{0}}{\mu_k + \sum_{j \in \mathcal{N}^{t+1}_{i+1}} \mu_j} \cr + &
            \frac{\sum_{j \in \mathcal{N}^{t+1}_{i+1}} \left( \frac{\mu_j}{\hat{\mu}^t_{i,j}} - \lambda_{i+1} a^{0}_{i+1, j}\right)}{\mu_k + \sum_{j \in \mathcal{N}^{t+1}_{i+1}} \mu_j}.
        \end{align}
    
        In other words, we have to check whether
        \begin{align}\nonumber
            C_i^t  <& 
            \left( \frac{\mu_k + \sum_{j \in \mathcal{N}^{t+1}_{i+1}} \mu_j}{\sum_{j \in \mathcal{N}^{t+1}_{i+1}} \mu_j}\right) \left(\frac{\lambda_{i+1} a^{0}_{i+1, h}}{\mu_h}\right) \cr+& 
            \lambda_{i+1} \left(\frac{1 - a_{i+1, k}^{0} -  \sum_{j \in \mathcal{N}^{t+1}_{i+1}}a^{0}_{i+1, j}}{\sum_{j \in \mathcal{N}^{t+1}_{i+1}} \mu_j}\right) \cr + &
            \frac{\sum_{j \in \mathcal{N}^{t+1}_{i+1}} \frac{\mu_j}{\hat{\mu}^t_{ij}}}{\sum_{j \in \mathcal{N}^{t+1}_{i+1}} \mu_j}.
        \end{align}
        More generally, suppose player $i+1$ sends jobs to servers in $\mathcal{H}_{i+1}^{t+1} \subseteq S_i^t$. Then for any $h\in S_i^t\setminus \mathcal{H}_{i+1}^{t+1}$ we have to check
        \begin{align}\nonumber
            C_i^t  <& 
            \left( \frac{\sum_{k \in \mathcal{H}_{i+1}^{t+1}} \mu_k + \sum_{j \in \mathcal{N}^{t+1}_{i+1}} \mu_j}{\sum_{j \in \mathcal{N}^{t+1}_{i+1}} \mu_j}\right) \left(\frac{\lambda_{i+1} a^{0}_{i+1, h}}{\mu_h}\right) \cr+& \lambda_{i+1} \left(\frac{1 - \sum_{k \in \mathcal{H}_{i+1}^{t+1}}a_{i+1, k}^{0} - \sum_{j \in \mathcal{N}^{t+1}_{i+1}}  a^{0}_{i+1, j}}{\sum_{j \in \mathcal{N}^{t+1}_{i+1}} \mu_j}\right) \cr +& \frac{\sum_{j \in \mathcal{N}^{t+1}_{i+1}} \frac{\mu_j}{\hat{\mu}^t_{ij}}}{\sum_{j \in \mathcal{N}^{t+1}_{i+1}} \mu_j},
        \end{align}
        which holds because of Eq.~\eqref{C_less_sum_mu} and the fact that $a^{0}_{i+1,j} \neq 0$ for all $j \in [m]$. Since $i \in \{1, \cdots, n-1\}$ is chosen arbitrarily, we can conclude that $S^1_{1}\subseteq S^{2}_2 \cdots \subseteq S^{n}_n$.
    \end{itemize}
    To complete the proof, we note that when player $n$ updates its action, it plays its best response with respect to the action profile $\{a_1^1, a_2^2, \cdots, a_{n-1}^{n-1}\}$. Suppose it sends jobs only to servers indexed in $\{1, \cdots c_n-1\}$. We therefore know that $S_i^i \subseteq \{1, \cdots c_n-1\}$ for $i \in \{1, \cdots n\}$. Moreover, we know that player $n$ distributes its job in a way such that
    \begin{align}
    \label{eq:equal-load-in-NE}
        C^n_n = \frac{\sum_{i=1}^n \lambda_{i} a^i_{ij} + s_j^0}{\mu_j} \ \ \forall j \in \{1, \cdots c_n-1\}.
    \end{align}
    In other words, player \( n \) ensures that the normalized load is equal across all servers receiving jobs. Consequently, for all players, the servers to which they allocate jobs have identical normalized loads, which is exactly what players achieve when playing their best-response strategies. As a result, no player has an incentive to change their policy, as this distribution aligns with their best-response strategy. Thus, \( \{a_1^1, a_2^2, \cdots, a_n^n\} \) forms a pure NE of the static game.$\hfill{\blacksquare}$
\end{pf}
\subsection{Proof of Theorem~\ref{thm:PoA}}\label{PoA-nonzero}

In order to prove Theorem \ref{thm:PoA}, we first state and prove the following auxiliary lemma. 

\begin{lem}\label{lemma:equal_load_NE}
Given an action profile \( a \), we define the \emph{normalized load} on server \( j \) as
\begin{align}\label{eq:L_j}
    L_j(a) \coloneqq \frac{s_j^0 + \sum_{i=1}^n \lambda_i a_{ij}}{\mu_j}.
\end{align}
Let \( a \) be a pure NE of the static load-balancing game, and denote its support by \( S(a) = \{j : a_{ij} > 0 \text{ for some } i\} \), i.e., the set of servers that receive jobs from at least one player under that NE. Then, the normalized loads on all servers in \( S(a) \) must be equal, that is,
\[
L_j(a) = L := \frac{\sum_{\ell \in S(a)} s_{\ell}^0 + \sum_{i=1}^n \lambda_i}{\sum_{\ell \in S(a)} \mu_{\ell}} \quad \forall j \in S(a).
\]
\end{lem}
\begin{pf}
In any pure NE of the static game, each player plays a best response \cite{nash1950equilibrium}, as characterized in Lemma~\ref{br-static}. Suppose $a$ is a pure NE. From Eq.~\eqref{bestResponsePolicyStatic}, for each server $j$ to which player $i$ assigns jobs, we have
\begin{align}\label{equal_load_NE}
    L_j(a)&=\frac{\lambda_i a_{ij}}{\mu_j}
    + \frac{s^0_j + \sum_{k \neq i} \lambda_k a_{kj}}{\mu_j}\cr
    &= \frac{\lambda_i + \sum_{k=1}^{c_i - 1} \tfrac{\mu_k}{\hat{\mu}_{ik}}}{\sum_{k=1}^{c_i - 1} \mu_k},
\end{align}
where we note that the right-hand side does not depend on $j$.  
Therefore, in an NE, whenever a player distributes jobs across multiple servers, it equalizes the normalized loads on those servers. 


Now suppose in an NE $a$ there exist two players $i$ and $i'$, where player $i$ distributes its job only to servers in a set $H$, while player $i'$ distributes its job only to servers in a disjoint set $H'$.\footnote{Otherwise, if $H \cap H' \neq \emptyset$, then by the first part of the proof, the normalized loads on all servers in $H \cup H'$ must be the same, and there is nothing to prove.} From Eq.~\eqref{equal_load_NE}, we know that in any NE, the normalized loads are equal across all servers within the same set. That is,
\[
\forall j \in H,\; L_j(a) = L \quad \text{and} \quad \forall j \in H',\; L_j(a) = L'.
\]
Our goal is to show that $L = L'$. To derive a contradiction, assume without loss of generality that $L' < L$. Since player $i$ does not send jobs to any server $j \in H'$, using Eq.~\eqref{def:c_i}, we obtain:
\begin{align*}
\frac{\lambda_i + \sum_{k \in H} \frac{\mu_k}{\hat{\mu}_{ik}}}{\sum_{k \in H} \mu_k} 
\leq\frac{1}{\hat{\mu}_{ij}}&= \frac{s_j^0 + \sum_{h \neq i} \lambda_h a_{hj}}{\mu_j} \cr
&=\frac{s_j^0 + \sum_{h} \lambda_h a_{hj}}{\mu_j}= L',  
\end{align*}
where the second equality holds because $a_{ij}=0$. On the other hand, we can expand the left-hand side as
\begin{align*}
    \frac{\lambda_i + \sum_{k \in H} \frac{\mu_k}{\hat{\mu}_{ik}}}{\sum_{k \in H} \mu_k} 
    &= \frac{\lambda_i + \sum_{k \in H}\left(s_k^0 + \sum_{h \neq i} \lambda_h a_{hk}\right)}{\sum_{k \in H}\mu_k} \\
    &= \frac{\sum_{k \in H}\left(s_k^0 + \sum_{h} \lambda_h a_{hk}\right)}{\sum_{k \in H}\mu_k} \\
    &= \frac{\sum_{k \in H} \mu_kL}{\sum_{k \in H}\mu_k}= L,
\end{align*} 
where the second equality holds because $\lambda_i=\sum_{k\in H}\lambda_i a_{ik}$. Thus, we have $L \leq L'$, which contradicts the assumption $L' < L$. 
Therefore, we must have $L = L'$.

Finally, using the above argument, since the definition of normalized load in \eqref{eq:L_j} is independent of a specific player $i$, we conclude that the normalized loads on all servers that receive jobs from at least one of the players must be equal, that is $L_j(a) = L\ \forall j\in S(a)$.
Therefore,
    \begin{align*}
        L\sum_{j\in S(a)} \mu_j=\sum_{j\in S(a)} \mu_j L_j(a)
        =\sum_{j\in S(a)} \left(s_j^0 + \sum_{i=1}^n \lambda_i a_{ij}\right).
    \end{align*}
    Using this relation and the identity $\sum_{i=1}^n\sum_{j\in S(a)} \lambda_i a_{ij} = \sum_{i=1}^n \lambda_i$, we obtain $L = \frac{\sum_{j\in S(a)} s_j^0+ \sum_{i=1}^n \lambda_i}{\sum_{j\in S(a)} \mu_j}$. $\hfill{\blacksquare}$
\end{pf}
\begin{pf}
    For an action profile $a$, the social cost equals
    \begin{align*}
        SC(a) &= \sum_{i=1}^n D_i(a) \\
        &= \sum_{i=1}^n \sum_{j=1}^m \lambda_i a_{ij} \left( \frac{\lambda_i a_{ij}}{2\mu_j} + \frac{s^0_j + \sum_{k \neq i} \lambda_k a_{kj}}{\mu_j} \right).
    \end{align*}
    Let $a'$ be any arbitrary pure Nash equilibrium, and let $a^*$ be the action profile that minimizes the social cost. Also define $NE := SC(a')$ and $OPT := SC(a^*)$. As in \cite{etesami2020smart}, for any $i \in [n]$, we can write 
    \begin{align*}
        D_i(a'_i, a'_{-i}) & \leq D_i(a^*_i, a'_{-i}) \\  & =
        \sum_{j=1}^m \lambda_i a^*_{ij} \left( \frac{\lambda_i a^*_{ij}}{2\mu_j} + \frac{s^0_j + \sum_{k \neq i} \lambda_k a'_{kj}}{\mu_j} \right).
    \end{align*}
    Summing the above inequality for all $i \in [n]$ we get:
    \begin{align}\nonumber
        NE &= \sum_{i=1}^n D_i(a'_i, a'_{-i}) 
        \\ & \leq\label{PoA:first-ineq}
        OPT + \sum_{i=1}^n \sum_{j=1}^m \lambda_i a^*_{ij} \left( \frac{s_j^0+\sum_{k =1}^n \lambda_k a'_{kj}}{\mu_j}\right). 
    \end{align}
    Using the result of Lemma~\ref{lemma:equal_load_NE}, we obtain:
    {\small\begin{align*}
        NE & \leq OPT + \sum_{i=1}^n \sum_{j=1}^m \lambda_i a^*_{ij} \left( \frac{s_j^0+\sum_{k =1}^n \lambda_k a'_{kj}}{\mu_j}\right)
        \\ & =
        OPT + L\sum_{i=1}^n \sum_{j\in S(a')} \lambda_i a^*_{ij}+\sum_{i=1}^n \sum_{j\notin S(a')} \lambda_i a^*_{ij} \left( \frac{s_j^0}{\mu_j}\right)\\ &\leq
        OPT + \max\{L, \max_{j\in[m]}\{ \frac{s_j^0}{\mu_j}\}\} \left(\sum_{i=1}^n \sum_{j=1}^m \lambda_i a^*_{ij}\right),
    \end{align*}}
    where $L = \frac{\sum_{j\in S(a')} s_j^0+\sum_{i=1}^n \lambda_i}{\sum_{j\in S(a')} \mu_j}$. Therefore, we have
\begin{align}\label{eq:PoA_original_eq}
        \frac{NE}{OPT} \leq 1 + \frac{\max\{L, \max_{j\in[m]}\{ \frac{s_j^0}{\mu_j}\}\} \left(\sum_{i=1}^n \lambda_i\right)}{OPT}.
    \end{align}
    Let us define $X_j(a):=\sum_{i=1}^n\lambda_i a_{ij}$. Then, we can rewrite the social cost of action profile $a$ as
    \begin{align}\nonumber
    SC(a) &= \sum_{j=1}^m \frac{1}{\mu_j}\left(\frac{X^2_j(a)}{2}+s_j^0X_j(a)\right),
    \end{align}
    which is a function of $X=(X_j)_{j \in [m]}$, such that \(X \in \mathcal{X}\), where $\mathcal{X} = \{ X \in \mathbb{R}_{\geq 0}^m : \sum_{j=1}^m X_j = \sum_{i=1}^n \lambda_i \}$.
    Thus,
    \begin{align*}
        OPT 
        &= \min_{X\in\mathcal{X}} \sum_{j=1}^m \frac{1}{\mu_j}\Bigl(\frac{X_j^2}{2}+s_j^0 X_j\Bigr)\ge \min_{X\in\mathcal{X}} \sum_{j=1}^m \frac{X_j^2}{2\mu_j} \cr
        & \ge \min_{X\in\mathcal{X}} \frac{\left(\sum_{j=1}^m X_j\right)^2}{2\sum_{j=1}^m \mu_j} 
        = \frac{\left(\sum_{i=1}^n \lambda_i\right)^2}{2\sum_{j=1}^m \mu_j},
    \end{align*}
    where the last inequality uses the Cauchy–Schwarz inequality. By substituting the above inequality in \eqref{eq:PoA_original_eq} we get
    \begin{align}\label{eq:better-PoA}
        \frac{NE}{OPT}\leq 1 + 2 \max\{L, \max_{j\in[m]}\{ \frac{s_j^0}{\mu_j}\}\}\left(\frac{\sum_{j=1}^m \mu_j}{\sum_{i=1}^n \lambda_i}\right).
    \end{align}
    Finally, we note that 
    \begin{align}\label{eq:L-final}
    L = \frac{\sum_{j\in S(a')} s_j^0+\sum_{i=1}^n \lambda_i}{\sum_{j\in S(a')} \mu_j}\leq \max_{j\in[m]}\{ \frac{s_j^0}{\mu_j}\} + \frac{\sum_{i=1}^n \lambda_i}{\mu_{\text{min}}}.
    \end{align}
    By substituting \eqref{eq:L-final} into \eqref{eq:better-PoA}, we obtain
    \begin{align*}
        \frac{NE}{OPT}\leq 1 + 2 \left(\max_{j\in[m]}\{ \frac{s_j^0}{\mu_j}\} + \frac{\sum_{i=1}^n \lambda_i}{\mu_{\text{min}}}\right)\left(\frac{\sum_{j=1}^m \mu_j}{\sum_{i=1}^n \lambda_i}\right).
\end{align*}$\hfill{\blacksquare}$
    \end{pf}

\begin{corollary}\label{corr:PoA-special}Suppose $s^0_j=0\ \forall j$. Then, $PoA\leq 3$.
\end{corollary}
\begin{pf}
In the special case where $s_j^0 = 0$ for all $j$, the support of any Nash equilibrium includes all servers, i.e., $S(a) = m$. In this case, the upper bound in Eq.~\eqref{eq:L-final} becomes $L = (\sum_{i=1}^n \lambda_i)/(\sum_{j=1}^m \mu_j)$. Substituting this into Eq.~\eqref{eq:better-PoA}, we obtain
\[
\text{PoA} \leq 1 + 2\left(\frac{\sum_{i=1}^n \lambda_i}{\sum_{j=1}^m \mu_j}\right) \left(\frac{\sum_{j=1}^m \mu_j}{\sum_{i=1}^n \lambda_i}\right) = 3.
\]  
\end{pf}
\subsection{An Alternative Bound for the Convergence Rate of the Dynamic Game}\label{appx:alternative-bound}

 In this appendix, we provide an alternative bound for the convergence rate of Algorithm~\ref{alg:main} in the dynamic load-balancing game, which, for certain ranges of game parameters, could be better than the convergence rate given in Theorem~\ref{dynamic-convergence-NE}. 
 
 As before, the main idea is to analyze how quickly $\sum_{j=1}^m s_j^t$ approaches zero. Define $\lambda_{\text{min}} =  \min_i \lambda_i$ and $\lambda_{\text{max}} =  \max_i \lambda_i$. Similarly, let $\mu_{\text{min}}=\min_j \mu_j$ and $\mu_{\text{max}}=\max_j \mu_j$. To lower-bound the amount of decrease in $\left|\sum_{j=1}^m s_j^{t+1} - \sum_{j=1}^m s_j^t\right|$, we consider two cases:
\begin{itemize}
    \item Not all servers are receiving jobs. In this case, the minimum change occurs when all servers, except the slowest one (the server with the lowest processing rate), receive and process jobs completely. For these servers, we have: $s_j^{t+1} - s_j^t = 0$. Thus, only the slowest server contributes to the change in $\sum_{j=1}^m s_j^t$. Without loss of generality, suppose the index of this server is $m$, i.e., $\mu_{\text{min}} = \mu_m$. There are two possible cases:
    \begin{itemize}
        \item[(1)] $\mu_{\text{min}} \leq s_m^t$: In this case we have $s_m^{t+1} = s_m^{t} - \mu_{\text{min}}$. This gives the inequality
        \begin{align*}
            \mu_{\text{min}} \leq \left| \sum_{j=1}^m s_j^{t+1} - \sum_{j=1}^m s_j^t \right|.
        \end{align*}
        \item[(2)] $s_m^t < \mu_{\text{min}}$: In this case we have $s_m^{t+1} = 0$, thus $s_m^{t+1} - s_m^{t} = - s_m^{t}$. Moreover, since server $m$ has not received any job at time $t$, according to Eq.~\eqref{c_samllest_index}, we have $\frac{s_m^{t}}{\mu_{\text{min}}} \geq \frac{\lambda_{\text{min}}}{\sum_{j=1}^{m-1} \mu_j}$. This gives us 
        \begin{align*} \frac{\mu_{\text{min}}\lambda_{\text{min}}}{\sum_{j=1}^{m} \mu_j - \mu_{\text{min}}} \leq \left| \sum_{j=1}^m s_j^{t+1} - \sum_{j=1}^m s_j^t \right|.
        \end{align*}
        
    \end{itemize}
    \item All servers are receiving jobs, but the state has not yet converged to zero: According to Theorem~\ref{dynamic-convergence-NE}, in this case we have 
    \begin{align*}
        \sum_{j=1}^m s_j^{t} - \sum_{j=1}^m s_j^{t+1} = \sum_{j=1}^m \mu_j - \lambda_i,
    \end{align*}
    where $i$ is the player sending job at time $t$. Therefore, we have
    \begin{align*}
        \sum_{j=1}^m \mu_j - \lambda_{\text{max}} \leq \left| \sum_{j=1}^m s_j^{t+1} - \sum_{j=1}^m s_j^t \right|.
    \end{align*}
\end{itemize}
Thus, putting all of the above cases together, we conclude that the number of iterations until convergence is upper-bounded by:
\begin{align*}
    t''&\leq \frac{\sum_{j=1}^m s_j^0}{|\sum_{j=1}^m s_j^{t+1} - \sum_{j=1}^m s_j^t|} \cr
    &\leq \frac{\sum_{j=1}^m s_j^0}{\min\left\{\sum_{j=1}^m \mu_j - \lambda_{\max},\  \mu_{\min},\ \frac{\mu_{\text{min}}\lambda_{\min}}{\sum_{j=1}^{m} \mu_j - \mu_{\min}}\right\}}.
\end{align*}
\end{document}